# 3D printed mesoporous superconductors with periodic order on three length scales and enhanced properties via block copolymer directed self-assembly

**Authors:** Fei Yu[1,2], R. Paxton Thedford[1,3], Thomas A. Tartaglia[1], Sejal S. Sheth[1], Guillaume Freychet[4,5], William R. T. Tait[1,3], Peter A. Beaucage[6], William L. Moore[1], Yuanzhi Li[7], Jörg G. Werner[7,8], Julia Thom-Levy[9], Sol M. Gruner[9], R. Bruce van Dover[1], Ulrich B. Wiesner[1,10,11]*

**Affiliations:**

[1] Department of Materials Science and Engineering, Cornell University, Ithaca, New York 14853, United States.

[2] Department of Chemistry and Chemical Biology, Cornell University, Ithaca, New York 14853, United States.

[3] Robert Frederick Smith School of Chemical and Biomolecular Engineering, Cornell University, Ithaca, New York 14853, United States.

[4] National Synchrotron Light Source-II, Brookhaven National Laboratory, Upton, New York 11973, United States.

[5] University of Grenoble Alpes, CEA, Leti, F-38000 Grenoble, France.

[6] NIST Center for Neutron Research, National Institute of Standards and Technology, Gaithersburg, Maryland 20899, United States.

[7] Department of Mechanical Engineering, Boston University, Boston, Massachusetts 02215, United States.

[8] Division of Materials Science and Engineering, Boston University, Boston, Massachusetts 02215, United States.

[9] Department of Physics, Cornell University, Ithaca, New York 14853, United States.

[10] Department of Design Tech, Cornell University, Ithaca, New York 14853, United States.

[11] Kavli Institute at Cornell for Nanoscale Science, Cornell University, Ithaca, New York 14853, United States.

* Corresponding author. Email: <u>ubw1@cornell.edu</u>

**Abstract:**

Solution-based soft matter self-assembly (SA) promises unique materials properties from approaches including additive manufacturing/three-dimensional (3D) printing. We report direct ink writing derived, hierarchically porous transition metal nitride superconductors (SCs) and precursor oxides, structure-directed by Pluronics-family block copolymer (BCP) SA and heat treated in various environments. SCs with periodic lattices on three length scales show record nanoconfinement-induced upper critical field enhancements correlated with BCP molar mass, attaining values of 50 T for NbN and 8.1 T for non-optimized TiN samples, the first mapping of a tailorable SC property onto a macromolecular parameter. They reach surface areas above 120 m$^2$/g, the highest reported for compound SCs to date. Embedded printing enables the first BCP directed mesoporous non-self-supporting helical SCs. Results suggest that additive manufacturing may open pathways to mesoporous SCs with not only a variety of macroscopic form factors but enhanced properties from intrinsic, SA-derived mesostructures with substantial academic and technological promise.

Additive manufacturing / three-dimensional (3D) printing dramatically expands the accessible range of form factors and has been extensively used in the field of soft matter[1]. While new printing techniques continue to advance this field[2–5], challenges remain in the 3D printing of self-assembly (SA) directed functional crystalline inorganic nanomaterials. Such materials may combine atomic lattices, SA-based mesoscale lattices, and 3D printing induced macroscopic lattices. This requires exquisite structure control over three separate length scales to achieve the desired properties, which explains why reports on such studies remain scarce. This is particularly true for quantum materials like superconductors (SCs), for which solution-based soft matter SA promises a combination of cost-effective synthesis approaches and mesostructure-induced property enhancements[6]. Block copolymer (BCP) SA-derived mesostructures, a hallmark of soft condensed matter science, have recently gained attention in 3D printing due to their exquisite structure control over mesoscale lattice parameters typically across 50-500 Å[7,8]. To the best of our knowledge, however, 3D printed BCP SA-based mesoporous SCs with periodically ordered lattices on three different length scales have not been reported. In fact, despite first reports on 3D printed BCP SA-directed mesostructured amorphous silica[9,10], a large part of the periodic table of elements remains unexplored. Transition metals and their compounds (*e.g.*, oxides and nitrides) are attractive candidates, as first experiments using large molar mass BCPs to structure direct mesostructured niobium nitride (NbN)-type materials have shown modified macroscopic SC behavior[11,12]. Compared to silica, however, their 3D printing is complicated by the rapid kinetics of the associated sol-gel synthesis processes[13].

Here, we report transition metal compounds accessible through sol-gel chemistry[13,14] in the form of nitrides with periodic atomic rocksalt lattices and their precursor oxides structure-directed by low molar mass Pluronics-family BCPs into mesoscale lattices and 3D printed by direct ink writing (DIW, Fig. 1) into various form factors including periodic cubic woodpile lattices and periodic helices. Identifying appropriate 3D printing and thermal processing conditions in various environments enables the successful formation of hierarchically porous nitride SCs with self-assembled mesostructures and record high surface areas. For these materials, to the best of our knowledge, we demonstrate the highest nanoconfinement induced enhancements to date for the upper critical field of the SCs. This enhancement is controllable by BCP molar mass, thereby mapping a SC property onto a macromolecular characteristic. In addition, we show the first 3D printed mesoporous superconducting helical structures, a non-self-supporting configuration particularly challenging to print, adding to the variety of achievable form factors. These results pave the way to classes of 3D printed SCs with substantial academic as well as technological promise.

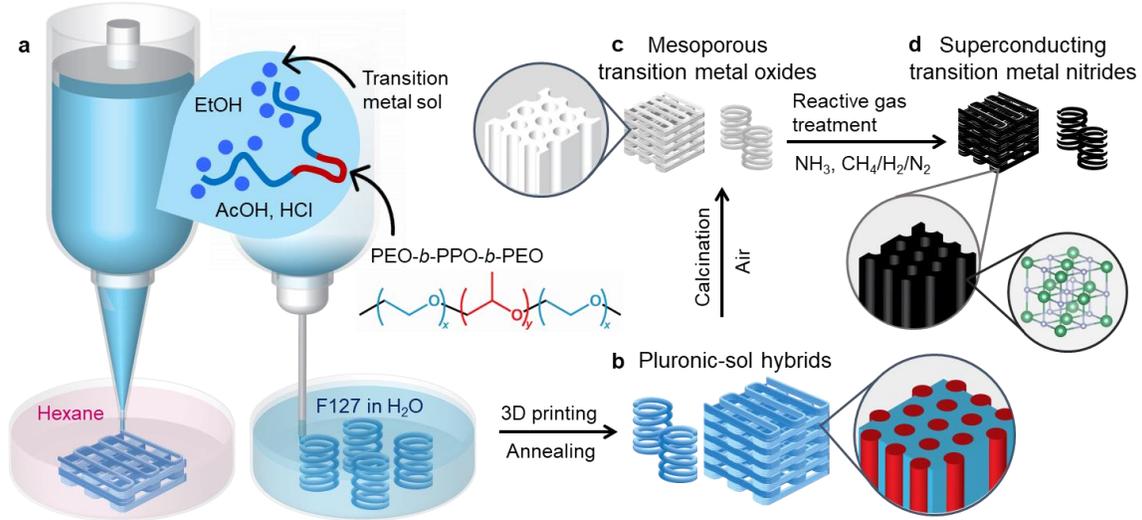

**Fig. 1. Schematic of the processes to achieve direct ink writing (DIW) derived and Pluronics block copolymer (BCP) self-assembly (SA) directed mesoporous superconductors (SCs) with periodic structures on three different length scales.** (**a**) The ink is composed of Pluronics-family BCPs mixed with transition metal sols hydrolyzed from metal alkoxides in acidic ethanol solutions. The syringe pump-type print head extrudes the ink into a dish containing either hexane for periodic cubic woodpile structures or gel-like 25 % Pluronic F127 by mass in water for periodic helical structures. After drying and annealing, (**b**) 3D printed Pluronics-sol hybrid woodpiles with self-assembled periodic hexagonal mesostructures and helices are calcined in air to yield (**c**) mesoporous transition metal oxides. After further heat treatments in ammonia and carburizing gas (mixture of methane, hydrogen, and nitrogen) at higher temperatures up to 950 °C, oxides are converted to (**d**) mesoporous superconducting transition metal nitride helices and hexagonally ordered woodpiles with cubic rocksalt atomic lattices.

**DIW derived and BCP SA directed porous superconductors with periodic lattices on three separate length scales**

We first selected amphiphilic Pluronics-family PEO-*b*-PPO-*b*-PEO ABA-type BCP F127 as the structure-directing agent for an inorganic niobia sol (Fig. 1). Directly writing the F127-niobia ink in air produced poor results, however, with solvent evaporation being too slow to lead to ink mechanical properties conducive to layer-by-layer depositions without collapse of the printed structures (Fig. S1b). Rheological results confirmed a liquid-like ink with low moduli inappropriate for 3D printing (Fig. S1a).

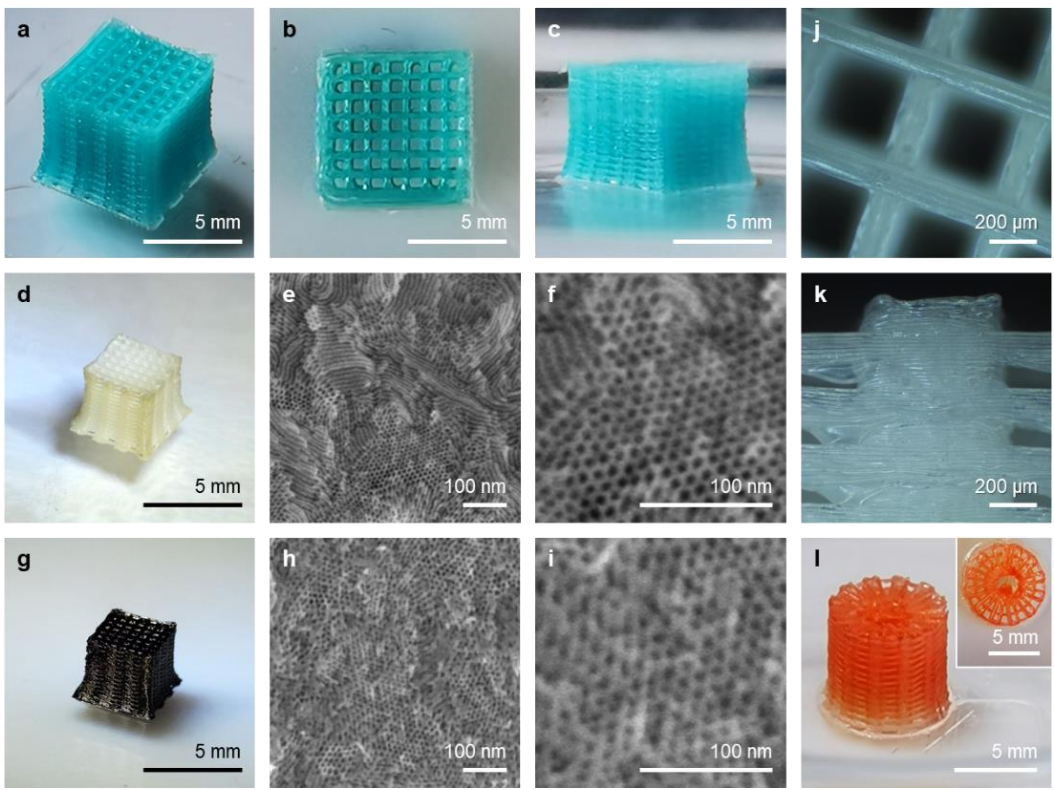

**Fig. 2. 3D printed structures with periodic atomic, mesoscale, and macroscopic lattices derived from BCP-niobia sol.** (**a**-**c**) Photos of as-printed periodic cubic woodpile of Pluronic F127-niobia sol hybrid. (**d**, **g**) Photos of resulting oxide (**d**) and nitride (**g**) woodpiles. (**e**, **f**, **h**, **i**) SEM micrographs at different magnifications of periodic hexagonal mesostructures of oxide (**e**, **f**) and nitride (**h**, **i**). (**j**, **k**) Optical micrographs of oxide woodpile in (**d**). (**l**) Photos of as-printed cylindrical woodpile of Pluronic F127-niobia sol hybrid. Blue (**a-c**) and orange (**l**) colors are from dyes dissolved in the ink.

Inorganic additives could in principle take on the role of rheological modifiers in such extrusion-based printing[4,15]. Targeting a particular morphology, *e.g.*, hexagonally packed cylinders, did not, however, leave much compositional leeway as the amount of inorganic sol relative to the BCP typically dictates the self-assembled morphology in the final BCP-additive hybrids[16]. To overcome this limitation, we improved ink printability by adopting a coagulation bath in which, upon extrusion, the polymer and/or additive would precipitate and gain the required mechanical strength for structure retention of printed parts[17]. Since the polyethylene oxide (PEO) blocks in the Pluronic BCPs and niobia sol are hydrophilic, we found a hydrophobic alkane bath such as hexane suitable to induce precipitation while preventing outward diffusion of the ink. In rheological tests, the ink storage modulus increased by more than two orders of magnitude upon immersion in hexane (Fig. S1b). Furthermore, storage and loss moduli quickly switched relative magnitudes with immersion time, leading to the desired more elastic response with increasing moduli. A 7 × 7 × 7 mm$^3$ periodic cubic woodpile could now be printed from such a DIW setup (Fig. 1 left side, Fig. S2, and movie S1), showing conformity to designed print paths and stable overhanging structures (Fig. 2a-c). Aside from a periodic cubic woodpile, other sample form factors could be printed as exemplified by a cylindrical woodpile (Fig. 2l).

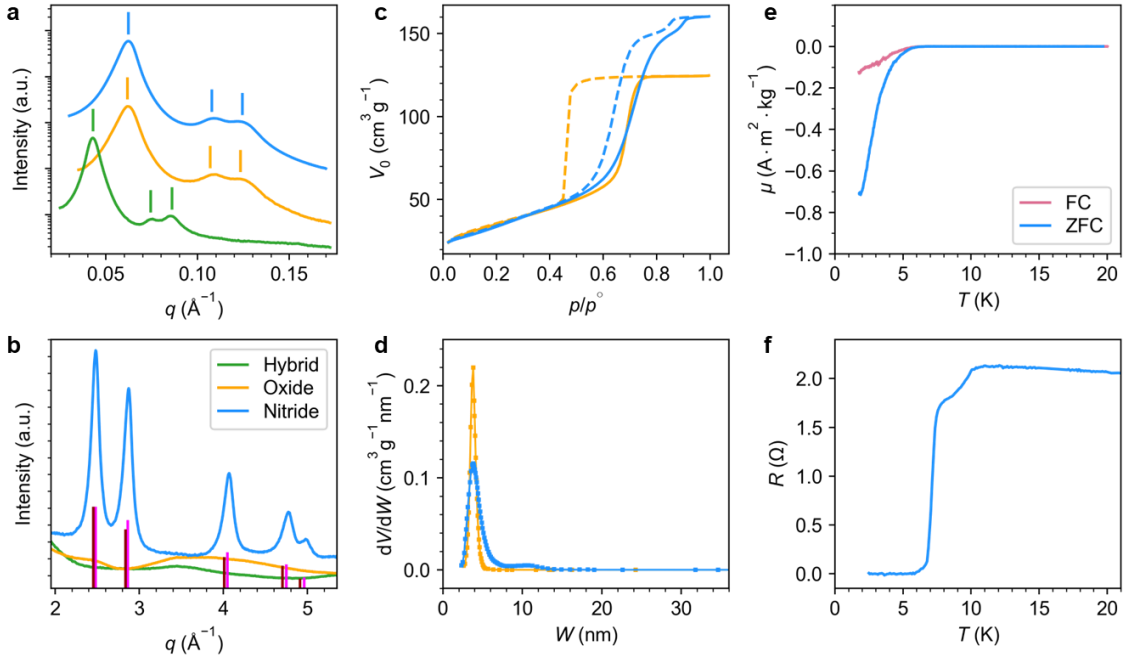

**Fig. 3. Characterization of mesoscale and atomic lattices as well as materials properties of 3D printed structures.** (**a**) Small-angle X-ray scattering (SAXS) and (**b**) wide-angle X-ray scattering (WAXS) profiles for 3D printed Pluronic F127-niobia sol hybrid and resulting oxide and nitride-type structures. Ticks in (**a**) denote relative scattering vector ($q$) positions (1:$\sqrt{3}$:2) of Bragg reflections for hexagonal structures (symmetry group P6mm). Dark red and magenta ticks in (**b**) represent relative intensities and $q$ positions of peaks of rocksalt δ-NbN (PDF #01-089-5007) and NbC (PDF #03-065-8781), respectively. They superimpose so closely that they are difficult to distinguish in (**b**). (**c**) Nitrogen adsorption (solid line) and desorption (dashed line) curves of mesoporous oxide and nitride. (**d**) Corresponding pore size distributions derived from the Barrett–Joyner–Halenda (BJH) model. $V_0$ is the volume of adsorbed nitrogen gas at standard temperature and pressure $p°$. $p/p°$, gas pressure $p$ divided by $p°$, is the relative pressure. $V$ is the pore volume and $W$ is the pore width. (**e**) Plot of magnetic moments normalized by sample mass

($\mu$) and (**f**) plot of electrical resistance (*R*) of nitride samples as a function of temperature (*T*). FC, field cooling; ZFC, zero-field cooling. Color codes in (**c-f**) are the same as in (**a, b**).

After thermal treatments at lower temperatures (40 to 60 °C) to remove residual solvents and further the condensation of the niobia sol, calcination in air at 450 °C decomposed the Pluronics BCP and resulted in white niobium oxide woodpiles with each dimension isotropically shrinking by about 30% relative to the hybrids (Fig. 2d,j,k). Thermogravimetric analysis showed the largest weight loss between 200 °C and 300 °C (Fig. S3). Locally, as revealed by the lack of well-defined diffraction peaks in wide-angle X-ray scattering (WAXS, Fig. 3b), these oxides (like the hybrids) are atomically amorphous. Scanning electron microscopy (SEM, Fig. 2e,f) suggested a hexagonally packed cylindrical pore structure on the mesoscale. This is corroborated by small-angle X-ray scattering (SAXS) results, showing Bragg reflections at relative *q* (scattering vector) positions of 1:$\sqrt{3}$:2 in reciprocal space (Fig. 3a), consistent with hexagonal lattices for hybrids and oxides. The observed shift of the primary peak position from 0.043 Å$^{-1}$ in the hybrid to 0.062 Å$^{-1}$ in the oxide corresponds to a 30 % reduction in mesoscopic periodicity, which is in agreement with the macroscopic sample shrinkage (compare panels in Fig. 2a and 2d).

The conversion of low molar mass Pluronics BCP SA-directed amorphous oxide materials with periodic mesoscale lattices to crystalline nitrides in general, and superconducting nitrides in particular, has remained a major challenge[18,19]. The low molar mass of these BCPs and associated thin inorganic walls favor collapse of the porous mesostructure during nitride crystal formation due to crystal overgrowth. Careful systematic studies allowed us to overcome this problem by identifying the following two-step chemical conversion approach (Supplementary Information): a first heating step under ammonia conducted at temperatures around 550 °C, *i.e.*, lower than suggested by earlier studies on large BCPs[11,12] to prevent mesostructure collapse but high enough for nitride crystal precipitation; then, without cooling the material to room temperature in-between as suggested by earlier studies[11,12], a second heating step to a higher temperature (750 to 950 °C) performed in carburizing gas ($CH_4$, $H_2$, and $N_2$ with molar ratios of 16:4:80) to improve nitride-type crystal quality but suppress grain growth. Relatively high flow rates of ammonia at 20 L/h and carburizing gas at 15 L/h (see SI for details) minimized the formation of crystalline oxide while simultaneously driving conversions to the desired final nitride-type materials.

After treating the white amorphous oxides in ammonia at 550 °C and in carburizing gas at 750 °C, woodpile structures turned reflective black (Fig. 2g). WAXS results on such 3D printed materials were consistent with phase-pure rocksalt δ-NbN (and/or NbC, Fig. 3b). Both macroscopic (Fig. 2g) and SAXS analysis based nanoscopic (Fig. 3a) structures and dimensions barely changed from those of the oxides. SEM inspection corroborated SAXS results, showing hexagonal mesostructures were preserved (Fig. 2h,i). Vibrating sample magnetometry (VSM) measured a critical temperature, $T_c$, of around 7 K, at which point the samples started to become diamagnetic (Fig. 3e), the prototypical superconducting behavior known as the Meissner effect. At around 7 K, the electrical resistance also dropped to zero (Fig. 3f), which confirmed that solution-processed and 3D printed samples were indeed SCs after heat treatments. Furthermore, nitrogen sorption/desorption results on oxides and nitrides exhibited type IV(a) isotherms with a hysteresis loop intermediate between H1 and H2(b), with pore sizes for both materials narrowly distributed around 4 nm according to the Barrett–Joyner–Halenda (BJH) model (Fig. 3c,d). For comparison, we also used non-local density functional theory (NLDFT) to estimate pore sizes, which typically came out larger than suggested by SEM (Fig. S4, Table S3). Brunauer–Emmett–Teller (BET) based data analysis yielded specific surface areas around 126 $m^2$/g and 120 $m^2$/g for oxides and nitrides, respectively.

Similar sol-gel synthesis was applied to the Pluronic-titania sol system to check how much these approaches may be generalizable (Fig. 1). Non-optimized first results using a two-step heating protocol demonstrated mesoporous superconducting TiN and its oxides (Supplementary Information, Figs. S5, S6).

**Controlling BCP SA directed superconductor properties**

After establishing access to 3D printed and BCP SA directed porous SCs, we explored the ability to control SC properties using the Pluronic-niobia sol system. Heat treatments of oxide samples to 750 °C in the second step were at the lower end of temperatures necessary for achieving superconductivity, consistent with material heterogeneities as suggested, *e.g.*, by kinks in the resistance curves (Fig. 3f). We therefore first carefully explored heating parameters to improve superconducting properties.

To that end, we varied the first (under $NH_3$) and second (under carburizing gas) heating step temperatures from 500 °C to 600 °C and 750 °C to 950 °C, respectively. For these ranges, we did not observe structure collapse, presumably because once nitrides are formed via the first-step treatment under ammonia, the temperature in the second-step treatment remains far below the nitride melting point ($T_m$(NbN) > 2,500 °C), minimizing Ostwald ripening[12]. The onset $T_c$ increased from around 7 K for samples treated first to 550 °C and then to 750 °C to 8-9 K and 15.2 K for samples treated to 850 °C and 950 °C in the second step, respectively, with concomitant monotonic increase in saturated magnetic moment normalized by sample mass (Fig. 4a). From the slope of field-dependent magnetic moment measurements (Fig. S7), a magnetic flux exclusion of ≈ 20 % was estimated for the sample treated to 950 °C as compared to a dense NbN SC with the same external dimensions (Supplementary Information). SEM micrographs of samples treated up to 950 °C retained hexagonally ordered mesostructures (Fig. 4a, inset), which was corroborated by SAXS data analyses. The periodic center-to-center distance between neighboring pores as calculated from SAXS results decreased from 11.7 nm, through 10.6 nm, to 10.3 nm for samples treated to 750 °C, 850 °C, and 950 °C, respectively (Fig. S8a). With increasing temperature in the carburizing gas treatment from 750 °C to 950 °C, the single-phase cubic unit cell sizes increased from 4.352 Å to 4.400 Å (Fig. S8b), while the coherently scattering domain sizes derived from Scherrer analyses of these data increased only slightly, from 4.0 nm to 4.7 nm, corroborating that substantial growth of crystalline domains was largely suppressed.

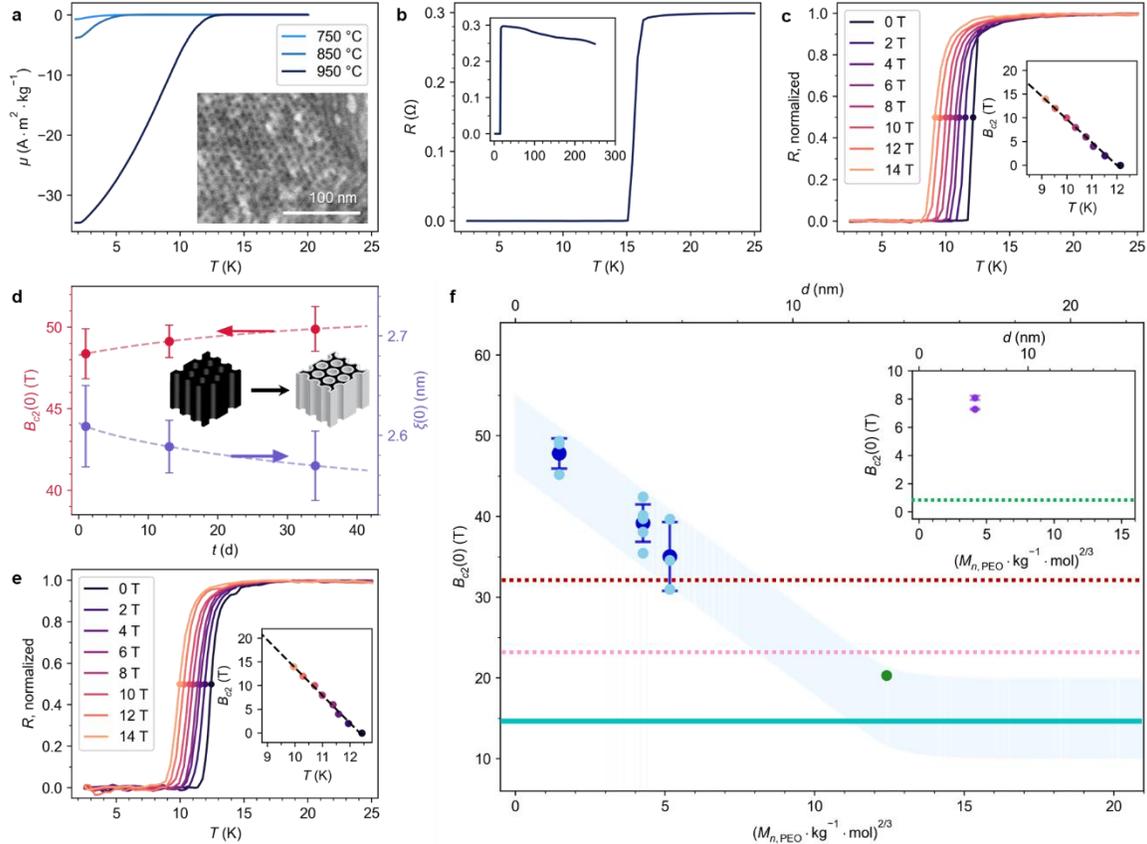

**Fig. 4. BCP directed superconductor (SC) properties.** (**a**) Plot of magnetic moment normalized by mass ($\mu$) of NbN-type samples treated to 550 °C in $NH_3$ and higher temperatures in carburizing gas (inset: SEM micrograph of sample treated to 950 °C). (**b**) Plot of electrical resistance ($R$) of NbN-type sample treated to 500 °C in $NH_3$ and 950 °C in carburizing gas. (**c**) Plot of field-dependent $R$ of sample in (**a**) treated to 950 °C (inset: variation of upper critical field ($B_{c2}$) with $T$ near $T_c$). (**d**) Plot of extrapolated $B_{c2}$ at $T = 0$ ($B_{c2}(0)$) (red) and corresponding Ginzburg–Landau coherence length ($\xi(0)$) (blue) as a function of room-temperature aging time ($t$, in days) in air for NbN-type sample treated to 575 °C in $NH_3$ and 950 °C in carburizing gas (error bars: extrapolation error; inset: schematic of aging). (**e**) Plot of field-dependent $R$ for sample in (**d**) aged 34 days. (**f**) Plot of $B_{c2}(0)$ for samples treated as in (**c**) as a function of $M_{n,PEO}^{2/3}$, proportional to wall

thickness, $d$ (top axis). For a given $M_{n,\text{PEO}}$, lighter blue points represent results from separate samples, while the dark blue point represents the average (error bars: standard deviation). The green point represents a hexagonally ordered sample derived from a tailor-made PI-*b*-PS-*b*-PEO. The light blue background is a visual guide. The teal line indicates the $B_{c2}(0)$ value for bulk NbN averaged from refs. 20–22. The dotted pink and red lines represent $B_{c2}(0)$ values of NbN in confined geometries from ref. 23 and ref. 24, respectively. The inset shows a plot of $B_{c2}(0)$ values (error bars: extrapolation error) for F127-directed TiN samples (Supporting Information, Fig. S19) against the dotted green line, which indicates the $B_{c2}(0)$ value reported for TiN under confinement from ref. 25.

Elemental compositions have been shown to influence NbN-type SC's behavior including critical temperatures (and fields, *vide infra*)[20,26]. In our synthetic approach, oxygen was anticipated to be present in the final materials, since nitridation in ammonia followed oxide formation in the first calcination step and nitrides are known to form surface oxide layers under ambient conditions, particularly relevant for the present high surface area materials (*vide infra*). Furthermore, we expected carbon in the final compositions, as methane in the carburizing gas treatment is expected to form radicals during decomposition at high temperatures and be reduced to carbonaceous species. X-ray photoelectron spectroscopy (XPS) of 3D printed samples first treated to 550 °C under ammonia and then to 750 °C, 850 °C, or 950 °C under carburizing gas, as well as after etching attempts to remove carbonaceous species deposited on pore surfaces (Supplementary Information, Table S1), suggested decreasing oxygen and increasing carbon content with increasing carburizing gas treatment temperature (Figs. S9 and S10, and Tables S1 and S2). We refer to these niobium oxycarbonitrides consistently as NbN-type materials throughout this text.

In contrast to typical metallic behavior, the resistance, $R$, of a 3D printed NbN-type material treated first to 500 °C and then to 950 °C increased upon cooling before reaching a $T_c$ above 15 K (Fig. 4b). This behavior could be ascribed to scattering occurring at grain boundaries within polycrystalline metallic conductors[27,28], consistent with the local granular structure revealed by high-resolution transmission electron microscopy (HRTEM) images of these materials (Fig. S11). This observed granularity in the BCP SA directed hexagonal mesostructure also hinted at a possibly shortened Ginzburg–Landau coherence length, $\xi$. For type-II SCs the upper critical magnetic field, $B_{c2}$, defined as $B_{c2} = \frac{\Phi_0}{2\pi\xi^2}$, where $\Phi_0 = \frac{h}{2e}$ is the magnetic flux quantum, $h$ the Planck constant, and $e$ the elementary charge, is inversely proportional to the square of $\xi$. Because of this high sensitivity to $\xi$, we investigated possible confinement induced enhancements in the upper critical field of our DIW derived and BCP directed SCs. According to WHH theory[29], the upper critical field at $T = 0$, $B_{c2}(0)$, can be extrapolated from the rate of change of $B_{c2}$ versus $T$ near $T_c$ via $B_{c2}(0) = 0.69 T_c \left. \dfrac{dB_{c2}}{dT} \right|_{T=T_c}$. Indeed, $T_c$ values obtained from field-dependent resistance measurements of a 3D printed material first treated to 550 °C and then to 950 °C extrapolated to a $B_{c2}(0)$ of 40 T (Fig. 4c). This is a substantial jump from bulk values[20–22] found between 10 and 20 T as well as the Pauli paramagnetic limit[30] of 1.84 T/K × $T_c$ ≈ 23 T. Elevated critical fields have been observed in SCs with confined dimensions, such as ultrathin films, monolayer materials, or ultrafine powders[23,24,31], but systematic studies of BCP SA based confinement effects on $B_{c2}$ in 3D printing derived mesostructured SCs remain unexplored.

To demonstrate that BCP SA based wall thickness induced confinement effects contribute to observed $B_{c2}(0)$ enhancements, we performed two control experiments. The first was to leave samples in the ambient air for extended periods of time allowing for surface oxide layer formation[32]. This should reduce nitride wall thickness of the porous BCP directed mesostructured SCs and therefore increase confinement effects. Aging in ambient air a NbN-type sample that had first been treated to 575 °C and then to 950 °C indeed increased the extrapolated $B_{c2}(0)$ (Fig. 4d) to a value of 50.1 T at the longest aging time of about a month tested (top red graph of Fig. 4d, and Fig. 4e). The extrapolated $B_{c2}(0)$ values obtained from aged samples also allowed estimates of the Ginzburg–Landau coherence length at $T = 0$, $\xi(0)$ (bottom blue graph of Fig. 4d). Considering materials heterogeneities, lines connecting extrapolated data points for $B_{c2}(0)$ and associated estimates for $\xi(0)$ were obtained from fits with a Kohlrausch–Williams–Watts (KWW) stretched exponential function with exponent, $\beta = 0.96$. Asymptotically reaching a $B_{c2}(0)$ value of 50.7 T obtained from this fit toward infinite time is consistent with self-limiting oxide surface layer formation. Estimated values for $\xi(0)$ around 2.6 nm are well below the single-crystal NbN value of $\xi_0(0) > 6$ nm[35]. As a result of the difference in pore sizes obtained from BJH and NLDFT models (Fig. 3d and Fig. S4b, respectively), the thinnest parts of the walls in the hexagonal mesostructures were estimated from SEM micrographs of a non-aged sample (Fig. 4a, inset) to be around 4.6 nm, *i.e.*, somewhat below the bulk $\xi(0)$ value for NbN. These estimates are all consistent with stronger mesostructural confinement associated with smaller physical nitride dimensions after oxidation leading to the observed increases in $B_{c2}(0)$. Diffuse electron scattering at wall surfaces will decrease the mean free path, $l_{tr}$ of electrons; in the dirty limit [$l_{tr} \ll \xi_0(T)$], Ginzburg-Landau theory predicts $\xi(T) = 0.855 [\xi_0(T) l_{tr}]^{1/2}$. Assuming $l_{tr}$ is comparable to the wall width (4.6 nm) we would expect $\xi(0)$ no smaller than roughly $0.855 \times (6 \text{ nm} \times 4.6 \text{ nm})^{1/2} = 4 \sim 5$ nm, *i.e.*, a value much greater than observed. We conclude that confinement introduces nontrivial effects on superconductivity in this system.

The second set of experiments to demonstrate BCP SA-based wall thickness-induced confinement effects was tuning mesostructure-associated wall thickness by varying BCP molar mass. To that end, in addition to F127 (≈ 12.6 kg/mol), alternative Pluronics-family

BCPs F108 (≈ 14.6 kg/mol) and P123 (≈ 5.8 kg/mol) with larger and smaller molar mass, respectively, than F127 were chosen for the preparation of mesoporous NbN-type SCs (Figs. S12 and S13). In order to deconvolute the contributions of thermal history and confinement effects on $B_{c2}(0)$, in these experiments all 3D printed and Pluronics BCP directed samples underwent identical 2-step thermal treatments to 550 °C under ammonia and then to 950 °C under carburizing gas (Figs. S14-S16). As a reference, a separate sample was prepared from a tailor-made poly(isoprene-*b*-styrene-*b*-ethylene oxide) (PI-*b*-PS-*b*-PEO) triblock terpolymer of a much larger molar mass (88.3 kg/mol), and then converted to the superconducting NbN-type materials (Fig. S17) employing established protocols (*i.e.*, without DIW)[12]. Extrapolated $B_{c2}(0)$ values as a function of the number average molar mass of the hydrophilic PEO block of the BCP (the block that mixes with the inorganic sols) raised to the power of 2/3, $M_{n,PEO}^{2/3}$, as a measure for its thickness, $d$ (top axis)[36], are summarized in Fig. 4f. For each of the Pluronics BCPs, at least 3 NbN-type samples (light blue) were tested and results averaged (points and associated error bars in dark blue). This plot reveals a clear trend: the shorter the structure directing BCP chains, the larger the critical field, $B_{c2}(0)$. The smallest BCP P123 resulted in the largest $B_{c2}(0)$ averaging near 50 T, while the largest customized BCP showed a $B_{c2}(0)$ close to bulk values.

Similar confinement effects were demonstrated for TiN. Direct nitridation of as-made hybrids in ammonia mitigated crystal coarsening and better retained the mesostructure directed by F127 compared with a two-step treatment via the oxide (Fig. S18). Non-optimized single-step heat treatment combined with aging protocols resulted in confinement induced $B_{c2}(0)$ enhancements up to values of 8.1 T from extrapolation (Fig. S19). This is a multifold increase over confinement induced values reported in the literature for TiN confined by varying film thickness, which are all under 1 T (Fig. 4f, inset)[25]. Results in Fig. 4d-f display, to the best of our knowledge, the highest nanoconfinement-induced $B_{c2}(0)$ enhancement to date achieved for NbN-type and TiN SCs that are correlated with BCP molar mass, the first mapping of a tailorable SC property onto a macromolecular control parameter[23,24,33,34]. Such solution-based BCP SA controlled and confinement-induced SC properties should be translatable to numerous macroscopic form factors via 3D printing and subsequent thermal conversion processes.

**3D printing complex non-self-supporting mesoporous structures**

One important application of high-$B_{c2}$ SCs is as coils in superconducting nuclear magnetic resonance (NMR) magnets. Brittle SCs are often challenging to bend into the coil-type shapes of electromagnets. What if in the future one could directly 3D print such complex non-self-supporting shapes? Compared with woodpiles, where printed layers underneath act as a support for building an overhanging feature above, non-self-supporting structures such as helices are challenging to form via DIW. Towards this goal and taking inspiration from embedded printing and other variants[37–39], a gel-like support matrix consisting of 25% F127 by mass in water was utilized to stabilize helical form factors printed from a F127-niobia sol ink (Fig. 1, movie S2). Instead of selectively crosslinking the ink or the support matrix to enable separation of the two, however, we worked in a pH neutral environment of F127 in water to induce rapid condensation of the niobia sol in the ink, which itself is at a strongly acidic pH to prevent premature aggregation of the sol nanoparticles. This inhibited substantial diffusion of the ink into the support matrix.

After cooling below the gel-to-fluid transition temperature (≈ 10 °C) for the F127 gel-like support, 3D printed periodic helices of various radii ranging from 2 mm to 4 mm and a pitch of 1.5 mm were removed and transferred to ethanol for rinsing (Fig. 5a). Helices in their hybrid state showed elastic behavior and could be reversibly compressed and released (Fig. 5b-d, movies S3 and S4). They could be scooped out of ethanol by tweezers for drying and subsequent heat treatments (Fig. 5e,f, movie S5). Freestanding inorganic oxide (Fig. 5g) and nitride-type (Fig. 5h) helices were obtained after calcination in air and the two-step heat treatments described earlier in ammonia and carburizing gas, respectively. While hybrids and oxides were again amorphous, WAXS profiles of the black nitride-type helices showed peaks matching those of rocksalt δ-NbN (and/or NbC, Fig. 5i).

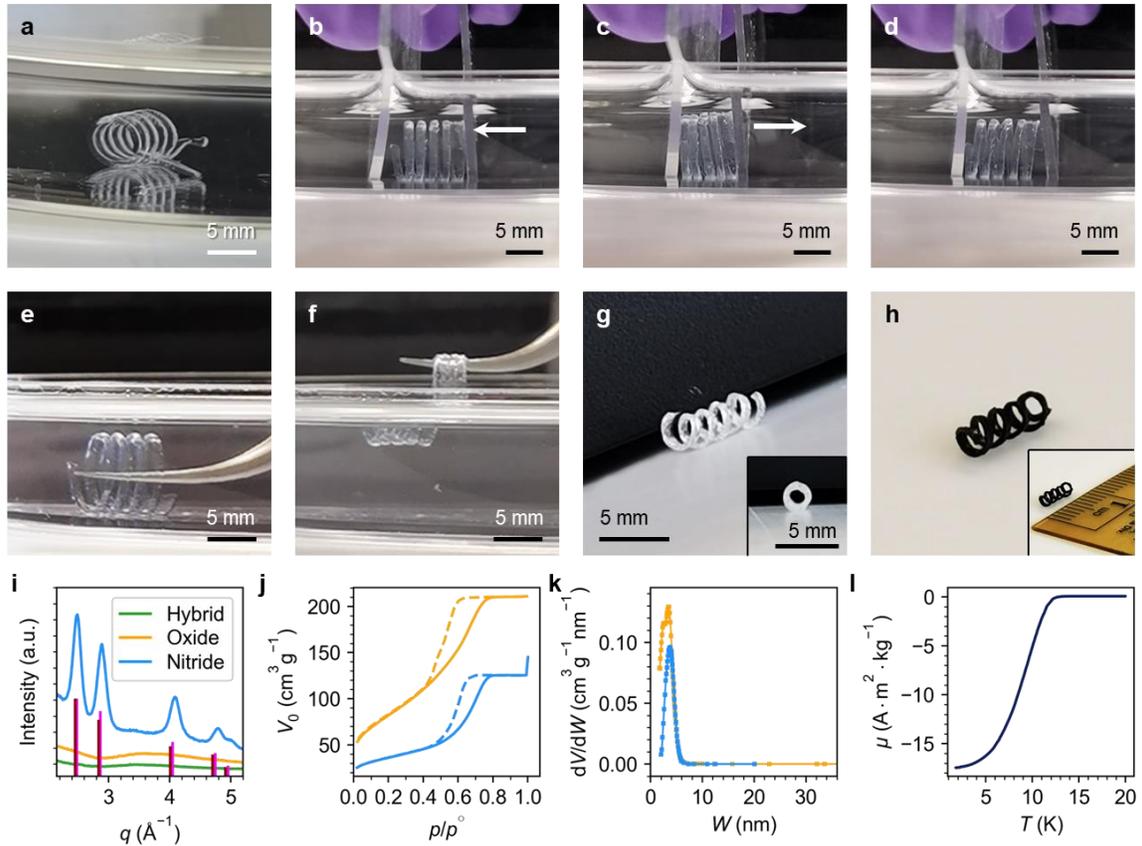

**Fig. 5. Superconducting periodic helical structures via embedded printing of F127-niobia ink.** (**a**) A helix in ethanol after removal from the support matrix used for embedded printing. (**b-d**) Visualizing the elastic behavior of a hybrid helix in ethanol. (**e, f**) Picking a hybrid helix out with tweezers (refraction causes apparent misalignment). (**g**) Oxide helix

after calcination in air to 450 °C (inset: view along helical axis). (**h**) Nitride helix after heat treatment to 950 °C in different environments (inset: helix dimension against a ruler). The helices in (**a**, **g**, and **h**) and (**b**-**f**) were printed with radii of 2 mm and 3 mm, respectively. (**i**) WAXS profiles of hybrid, oxide, and nitride helices (nitride heat treated in carburizing gas to 750 °C). Dark red and magenta ticks represent relative intensities and corresponding $q$ positions of peaks of NbN (PDF #01-089-5007) and NbC (PDF #03-065-8781), respectively. (**j**) Nitrogen adsorption (solid line) and desorption (dashed line) curves of mesoporous oxide and nitride helices, and (**k**) corresponding pore size distributions derived from the BJH model. Same color scheme in (**i**) applies to (**j**, **k**). (**l**) Magnetic moment normalized by mass ($\mu$) of nitride helix heat treated to 950 °C.

Resulting 3D printing derived and Pluronics BCP SA directed oxide and nitride-type periodic helical materials treated first to 550 °C under ammonia and then to 750 °C in carburizing gas were highly porous, reaching BET specific surface areas of 298 m$^2$/g and 129 m$^2$/g, respectively (Fig. 5j), with the latter even slightly surpassing the value for the similarly treated NbN-type woodpile structures described earlier. Narrow pore size distributions were centered around 4 nm for oxide and nitride-type materials both characteristic of F127-directed mesostructures (Fig. 5k). VSM measurements confirmed that a mesoporous helix treated at 950 °C was a SC with an onset $T_c$ of 13.5 K (Fig. 5l). To the best of our knowledge, the 3D printed and heat processed materials described herein (both woodpiles and helices) constitute the highest surface area mesoporous compound SCs, and the mesoporous helices the first non-self-supported 3D printed SCs reported to date[40,41].

**Conclusions**

Mesoporous SCs and precursor oxides with Pluronics BCP SA-directed mesostructures were fabricated by direct ink writing and subsequent thermal treatments in various gas environments. Either a hydrophobic precipitation bath or a gel-like support matrix helped maintain the integrity of complex 3D printed structures, including non-self-supporting helices. The work combined the freedom to choose various macroscopic periodic form factors by additive manufacturing with the periodic mesostructural control of BCP SA directed highly functional inorganic materials with periodic atomic lattices. Results suggest that solution-based 3D printing approaches enable generation of hierarchically ordered porous SCs with properties substantially enhanced over bulk behavior by nanoconfinement, and variable form factors for a range of applications including sensing and energy. Furthermore, such SCs accessible from commercially available chemicals, polymers, and 3D printing equipment should enable a plethora of advanced studies in various scientific communities, *e.g.*, linking the hitherto largely disconnected fields of additive manufacturing, mesoporous materials, and interfacial properties of correlated electron systems, and facilitate bridging fundamental studies with application-oriented "proof-of-principle" device developments.

**Acknowledgments:**

We would like to thank Louisa Smieska for experimental assistance at CHESS, and Patryk Wasik for experimental assistance at NSLS-II. We would like to thank John Grazul for help in TEM characterizations and John Wright for help with TGA and XPS experiments. We would also like to thank Danni Tang for help with photography.

This work was supported by the National Science Foundation (NSF) under grant DMR-2307013.

We acknowledge the use of facilities and instrumentation supported by NSF through the Cornell University Materials Research Science and Engineering Center DMR-1719875.

We also acknowledge the use of the SMI beamline of the National Synchrotron Light Source II (NSLS-II), a U.S. Department of Energy (DOE) Office of Science User Facility operated for the DOE Office of Science by Brookhaven National Laboratory under Contract No. DE-SC0012704, and the use of the FMB beamline of Cornell High Energy Synchrotron Source, sponsored by Air Force Research Laboratory (AFRL) under agreement numbers FA8650-19-2-5220 and FA8650-22-2-5200. Part of the SAXS experiments were performed at the Center for High-Energy X-ray Sciences (CHEXS), which is supported by the National Science Foundation (BIO, ENG and MPS Directorates) under award DMR-1829070, and the Macromolecular Diffraction at CHESS (MacCHESS) facility, which is supported by award 1-P30-GM124166-01A1 from the National Institute of General Medical Sciences, National Institutes of Health, and by New York State's Empire State Development Corporation (NYSTAR).

**Author contributions:**

Conceptualization: FY, RPT, UBW
Methodology & interpretation: FY, RPT, UBW
Investigation: FY, RPT, TAT, SSS, GF, WRTT, PAB, WLM, YL, JGW
Visualization: FY, UBW
Funding acquisition: UBW
Project administration: UBW
Supervision: UBW
Writing – original draft: FY, UBW
Writing – review & editing: FY, RPT, TAT, SSS, GF, WRTT, PAB, WLM, YL, JGW, JT, SMG, RBV, UBW

**Competing interests:** The authors declare that a patent disclosure has been filed with Cornell's Center for Technology Licensing (CTL).

**Data availabilities:** All data are available in the main text or the supplementary materials. The G-code files for controlling the print head movement are available upon request.

**Additional information:**

Methods
Supplementary Text
Figs. S1 to S25
Tables S1 to S4
Movies S1 to S5
Glossary
Supplementary References

Supplementary Information for

**3D printed mesoporous superconductors with periodic order on three length scales and enhanced properties via block copolymer directed self-assembly**

Fei Yu, R. Paxton Thedford, Thomas A. Tartaglia, Sejal S. Sheth, Guillaume Freychet, William R. T. Tait, Peter A. Beaucage, William L. Moore, Yuanzhi Li, Jörg G. Werner, Julia Thom-Levy, Sol M. Gruner, R. Bruce van Dover, Ulrich B. Wiesner*

* Corresponding author. Email: ubw1@cornell.edu

# 1 Methods

## 1.1 Ink preparation

0.40 g Pluronic F127 (Sigma-Aldrich) was mixed with 0.40 g anhydrous ethanol ($C_2H_5OH$, Sigma-Aldrich, ≥99.5 %), 0.185 mL 37 % hydrochloric acid (EMD Millipore), and 0.285 mL glacial acetic acid ($CH_3COOH$, Macron), and stirred until fully dissolved. For printing structures directed by other Pluronic block copolymers (BCPs), the F127 was replaced by the same amount of F108 (Alrich Chemistry) or P123 (BASF). For P123, F127, and F108, ($x$, $y$) subscripts in the formula $PEO_x$–$PPO_y$–$PEO_x$ are (20, 70), (100, 65), and (133, 50), respectively, *i.e.*, the PEO block size hybridizing with niobia sol nanoparticles decreases in the sequence F108, F127, and P123, while the PPO block size responsible for the pore size in the mesoporous materials decreases in the reverse sequence. While stirring vigorously, 0.625 mL niobium(V) ethoxide ($Nb(OC_2H_5)_5$, Alfa Aesar, 99.99 %, metal basis) was quickly injected into the F127 solution. For inks made with P123, up to 0.675 mL niobium ethoxide was added. The clear F127-niobia sol solution was left stirring for 3 days before being loaded into a 30 mL syringe (BD) with a tapered dispensing tip (Nordson, 27 gauge), or a blunt needle (CML Supply, 25 gauge, up to one and half inches long) for embedded printing. 50 μL of ethanol solutions of methylene blue (1 mg/mL) or Sudan I (saturated) could be added to generate a colored ink.

For inks containing titania sol, the materials and procedures were the same except that the 0.625 mL niobium(V) ethoxide was substituted with 0.74 mL titanium(IV) tetraisopropoxide ($Ti(OCH(CH_3)_2)_4$, 99.995 %, Thermo Scientific Chemicals).

## 1.2 Direct ink writing (DIW)

The ink containing syringe was mounted on an SDS30 print head connected to a Hyrel SR 3D printer (Fig. S20). The movement of the print head and the platform was controlled by the built-in Repetrel software (version 4.2.494) with customized G-codes. DIW occurred in a glass petri dish (printing dish, 60 mm diameter, at least 15 mm depth) filled with hexane (Fisher chemical) and periodically replenished by a syringe pump (Fig. S20) to keep printed parts submerged throughout the process. Each step of movement in the vertical direction was 30 μm.

For woodpile structures, the print head would move at a speed of 100 to 300 mm/min relative to the platform along the same path 8 times consecutively before switching directions to build enough thickness for each strut while allowing thin layers of ink to precipitate thoroughly in hexane. A slower speed of 100 mm/min was used for F127-titania sol inks, and the woodpile top during printing was kept just below the hexane surface so as to reliably produce high-quality printed structures. A typical cubic woodpile measured 7 × 7 × 7 $mm^3$, with 1 mm between adjacent in-plane struts and a total of 240 layers. After printing was accomplished, hexane was poured out and a cover dish was put on the printing dish with the printed part still wet. In this configuration printed parts were aged in two steps before further heat treatments at higher temperatures to generate oxides and nitrides. In the first step, samples were brought to a slightly elevated temperature of 40 °C for one day. Samples were subsequently heat treated at 60 degrees in a second step for another day. After aging, residual solvents had evaporated and the BCP hybrid sol-gel materials consolidated.

For embedded printing, the printing dish was filled with a gel-like support matrix made from 25 % F127 by mass in water. On top of this gel-like matrix was deposited via pipette a ≈ 2 mm thick liquid layer of 21 % F127 by mass in water. This layer allowed backfilling of the crevices formed by the needle traversing through the support matrix during printing with F127 in water. In contrast to the multilayer approach used during printing of woodpile structures in hexane, there was no such multilayer approach taken for printing helices, *i.e.*, there was no repetition along the same path during the embedded printing. The radii of the helices as set in the G-codes varied from 2 to 4 mm and the pitch was 1.5 mm (Movie S2). The print head speed was 100 mm/min to 150 mm/min. After the printing process concluded, like for the DIW into hexane, samples were aged in two steps. In the first step, a cover dish was sealed by parafilm with the printing dish containing the matrix and the printed part, which was then kept at room temperature for 1 day. In the second step, the sealed dishes with the printed samples were heated at 60 °C for another day to further consolidate printed hybrid materials. After these two aging steps and before further heat treatments of the printed parts, the support matrix and the fluid filler were separated from the printed parts by cooling below the gel-to-fluid transition temperature (≈ 10 °C) so that the printed parts could be picked out and rinsed with ethanol (EMD Millipore, 96%) (Movie S5). Once residual ethanol had evaporated, parts were ready for further heat treatments at higher temperatures.

1.3 Heat treatments in different gas environments

The BCP-niobia hybrid printed parts were first heated at a ramp rate of 1 °C/min to 450 °C with a dwell time of 3 h in air in a tube furnace with an alumina tube of 1 inch diameter and ≈ 1 m length. To further improve structural integrity, a preceding heating step in nitrogen gas with the same ramp rate, dwell time, and dwell temperature could be used. The resulting niobium oxide parts were then heated in another tube furnace with a quartz tube of 1 inch diameter and ≈ 60 cm length under ammonia from room temperature (≈ 20 °C) to 500 °C to 600 °C in 100 min with a dwell time of 210 min, and subsequently followed by heating to the final highest temperature of 750 °C to 950 °C at a ramp rate of 20 °C/min with a dwell time of 90 min. Ammonia gas ($NH_3$, Phoenix Electronics, electronic grade) flowed through the tube furnace at a flow rate of 20 L/h until it was switched (after being held for 180 min at the intermediate, *e.g.*, 550 °C, temperature) to carburizing gas (methane, $CH_4$, hydrogen, $H_2$, and nitrogen, $N_2$, mixed in a molar ratio of 16:4:80, Airgas) at a flow rate of 15 L/h. The NbN-type samples were left to cool to room temperature while nitrogen gas flowed for 2 h to purge the tube. The nitrogen gas flow rate was lowered to 10 L/h at temperatures below 450 °C.

The BCP-titania hybrid printed parts were heated at a ramp rate of 1 °C/min to either 300 °C or to 400 °C with a dwell time of 3 h in air in the tube furnace with the alumina tube. 300 °C treated samples would result in amorphous titania while 400 °C treated samples would show the formation of crystalline anatase. The nitridation step took place all in ammonia using the same heating protocol as for niobia, using the amorphous oxide calcined in air at 300 °C. Superconducting TiN in Fig. 4f, however, was derived by heating the hybrid materials in ammonia directly, reaching 500 °C in 25 min from room temperature to reduce the time spent at lower temperatures thereby suppressing crystallization of the oxide. After holding at 500 °C for 3 h, the sample was further heated at a ramp rate of 20 °C/min to 900 °C or higher with a dwell time of 1 h before cooling to room temperature under ambient conditions. Ammonia flow rate was kept at 15 L/min throughout the process to drive the nitridation forward.

1.4 NbN-type samples prepared from poly(isoprene-*b*-styrene-*b*-ethylene oxide)

Procedures described in prior publications[11,12] were followed to prepare superconducting NbN-type samples structure-directed by poly(isoprene-*b*-styrene-*b*-ethylene oxide) (referred to as ISO). The ISO terpolymer was synthesized via sequential anionic polymerization reported elsewhere[42]. Proton nuclear magnetic resonance ($^1$H NMR) and gel permeation chromatography (GPC) were used to determine the total number-average molar mass ($M_n$) of 88.3 kg/mol and a polydispersity index (PDI, or using the updated term dispersity, $Đ$) of 1.14, with mass fractions of PI, PS, and PEO blocks being 13.4 %, 37.1 %, and 49.5 %, respectively. For BCP directed hybrid formation, 0.050 g of ISO was dissolved in 1.0 mL anhydrous tetrahydrofuran (THF, 99.9%, Sigma-Aldrich). In a separate vial to prepare the sol stock solution, 0.96 mL niobium (V) ethoxide was injected to a mixture of 0.56 mL 37 % HCl solution and 0.90 mL THF under vigorous stirring. After 5 min, another 4.5 mL THF was added followed by additional stirring of 2 min. Targeting the hexagonal morphology, 0.740 mL sol stock solution was added to the ISO solution, which was cast into a Teflon beaker after mixing thoroughly overnight and left to evaporate under a hemispherical glass dome at 50 °C. After evaporation-induced self-assembly the dried film was heated at 130 °C in a vacuum oven for 3 h before calcination in the tube furnace in the same way as the 3D printed samples. The reactive gas treatment took place at a higher temperature and for a longer duration to chemically convert the thicker walls of ISO-directed structures. Using the previous protocol, the oxide film was heated in ammonia for 9 h at 700 °C and in carburizing gas for 3 h at 1000 °C to prepare the final superconducting NbN-type material.

## 1.5 Characterization

### 1.5.1 Optical imaging

Optical images were captured by a Samsung Galaxy S8 phone, a Huawei P40 pro+ phone or an Andonstar microscope. Fig. S21 was obtained using a Zeiss SteREO Discovery.V12 optical microscope equipped with an Olympus OM-D E-M5 camera.

### 1.5.2 Scanning electron microscopy (SEM)

Pieces broken from printed parts (oxide or nitride) were imaged using a Zeiss Gemini500 scanning electron microscope with an in-lens EsB detector. The accelerating voltage was 2 kV and the working distance was around 4 mm. No conductive coating needed to be sputtered onto samples prior to imaging. SEM-based pore size and wall thickness estimates were obtained using the ImageJ software. Results were averaged over at least 10 measurements per estimate.

### 1.5.3 Transmission electron microscopy (TEM)

Pulverized NbN-type samples from printed parts were suspended in absolute methanol and vortex mixed. 5 µL of the suspension was placed on a lacey carbon supported copper TEM grid (size 200 mesh) and left to evaporate. Samples were imaged in a Spectra 300 TEM (Thermo Fisher Scientific) with a cold field emission gun at 300 kV.

### 1.5.4 X-ray scattering

Small-angle and wide-angle X-ray scattering (SAXS and WAXS) were performed on pieces broken from printed parts at the Soft Matter Interfaces (SMI, 12-ID) beamline at the National Synchrotron Light Source II (NSLS-II) or the Functional Materials Beamline (FMB, ID3B) and BioSAXS Beamline at the Cornell High Energy Synchrotron Source (CHESS). At SMI, the X-ray energy was 16.1 keV. The SAXS detector was a Pilatus 1M pixel array detector at a distance of 5.00 m from the sample. The WAXS detector was a Pilatus 900KW pixel array detector. Two-dimensional (2D) scattering signals collected at various detector positions were stitched together to construct the WAXS scattering profiles over the entire $q$ (scattering vector) range. At FMB, the X-ray energy was 9.79 keV. The SAXS detector was a Pilatus 300K pixel array detector at a distance of 1.43 m from the sample. The WAXS detectors were two Pilatus 200K pixel array detectors. At BioSAXS, the X-ray energy was 11.24 keV. The SAXS detector was an Eiger 4M detector at a distance of 1.78 m from the sample. All the 2D raw scattering data were processed and azimuthally integrated using custom-generated software packages at the respective beamlines. Crystal unit cell size and the coherently scattering domain size were obtained by performing whole pattern fitting (Rietveld refinement) against NbN (PDF #01-089-5007) using the JADE software (Fig. S8b, Tables S3 and S4).

### 1.5.5 Nitrogen sorption

Nitrogen sorption was conducted on a Micromeritics ASAP 2020 Accelerated Surface Area and Porosimetry System at 77 K. Printed and heat treated parts were degassed overnight at 120 °C prior to measurement. The surface area was obtained based on the Brunauer–Emmett–Teller (BET) theory. Pore size distributions were calculated using the non-local density functional theory (NLDFT) model or the Barrett–Joyner–Halenda (BJH) model provided by the associated instrument software. The NLDFT model assumed nitrogen in cylindrical pores with oxide surface. The adsorption curve was used for BJH model analysis with Faass correction. All nitrogen sorption data shown in this paper for NbN-type materials were obtained from samples heat treated to 750 °C in carburizing gas.

*1.5.6 Rheology*

Rheological ink sample characterization was performed on a DHR3 rheometer by TA Instruments using a cone-and-plate geometry at 25 °C. The diameter for both cone and plate was 40 mm. The cone angle was 2 ° and the truncation on the cone was 63 μm. A solvent trap enclosed the geometry during measurements to minimize effects from solvent evaporation. Apart from being directly measured, the ink was also separately coated on cone and plate parts and then immersed in hexane for various times (Fig. S22) before cone and plate assembly for rheological measurements. Three sets of measurements were performed. First, the storage and loss moduli were measured as the strain was swept from 0.01 % to 100 % at a frequency of 1 Hz in the oscillatory mode. Second, the storage and loss moduli were measured as the frequency was swept from 0.1 Hz to 100 Hz at 1 % strain in the oscillatory mode. Lastly, the viscosity was measured as the shear rate was swept from 0.01 to 100 $s^{-1}$ in the continuous flow mode.

*1.5.7 X-ray photoelectron spectroscopy (XPS)*

Samples were first pulverized and then analyzed using a Thermo Scientific Nexsa G2 Spectrometer with an operating pressure of approximately $1\times10^{-9}$ Torr. Monochromatic Al $K_\alpha$ X-rays (1486.6 eV) with photoelectrons were collected from a 400 μm diameter analysis spot at a 90° emission angle with source to analyzer angle of 54.7 °. A hemispherical analyzer determined electron kinetic energy, using a pass energy of 200 eV for wide/survey scans, and 50 eV for high-resolution scans. A flood gun was used for charge neutralization of non-conductive samples. Etching was performed using an argon ion gun with an accelerating voltage of 500 V and an etch time of 120 s. The estimated etch depth was 10 nm. Data were analyzed using the CasaXPS software. Atomic percentages were quantified based on the survey scans using default regions of different elements in the software. Chemical bonding information was obtained from high-resolution scans by fitting the C 1s peak with carbide, C=O, C–C, and C–O components with the same relative sensitivity factor (RSF) and a mixed Gaussian/Lorentzian line shape (70/30).

*1.5.8 Thermogravimetric analysis (TGA)*

Thermogravimetric analysis was performed on 3D printed F127-niobia hybrid by heating at a rate of 4 °C/min in air in a TA Instruments 5500 thermogravimetric analyzer.

*1.5.9 Vibrating sample magnetometry (VSM)*

A Quantum Design Dynacool Physical Property Measurement System (PPMS) was used to collect VSM data. The sample mass was determined on a Cahn 28 automatic electrobalance. A printed sample piece was loaded into a pair of plastic holders that were then pressed into brass holders. Samples were cooled down to 1.8 K under zero external field (zero-field cooling, or ZFC) or 100 or 1000 Oe external field (field cooling, or FC). The sample's magnetic moment was measured as the temperature was ramped to 20 K at a rate of 1 K/min. For field dependent magnetization measurements, the external field was ramped to 50 kOe, then to -50 kOe, and back to 0 at a rate of 100 Oe/s. Sample magnetic moments were measured with an averaging time of 1 s per measurement. For flux exclusion calculations, it was assumed that the mass of a 3D printed structure was proportional to the volume enclosed by the sample's outer surface. For example, the woodpile nitride shown in Fig. 2g of the main text has a mass of ≈ 50 mg with an external dimension of (7 × 0.7 mm)$^3$ = 117 mm$^3$ (30 % shrinkage in each dimension with respect to the original printed hybrid woodpile cube). Therefore, a sample of mass *m* has a volume $V = \frac{m}{50 \text{ mg}} \times 117 \text{ mm}^3$ enclosed by its outer surface. At any externally applied field strength *H*, an ideal dense NbN superconductor with this volume would have a magnetic moment *μ* of -*HV*, by which the experimentally observed *μ* of the sample is divided to obtain the percentage of flux exclusion.

*1.5.10 Electrical transport measurements*

Field-dependent electrical resistance was measured on a Quantum Design PPMS equipped with either a 9 T or 14 T magnet. A sample piece broken from a printed part was fixed using cryogenic varnish on a silicon substrate with a 100 nm wet thermal oxide finish. The substrate's four corners were pre-patterned with four 40 nm gold contact pads with 20 nm chromium as the adhesion layer. 503 silver conductive paint (Electron Microscopy Sciences) was used to make four contacts between the sample and the gold contact pads (Fig. S21).

The silicon substrate carrying the sample was mounted and bonded through aluminum wires onto a Quantum Design resistivity puck using a West Bond 747630E wire bonder. After loading the puck into the PPMS, low-temperature resistance values were collected from 2.5 K up to 25 K at a direct current of 100 µA for each magnetic field from 0 up to 14 T at an increment of 2 T. For TiN samples, the magnetic field ranged from 0 to 2 T in increments of 0.2 T. A full temperature range up to 305 K of resistance was also measured as the sample warmed up to room temperature. Each data point was averaged from 5 or 25 measurements. $T_c$ was determined as the midpoint of the resistance curve. The extrapolation error derived from the plots of $B_{c2}$ versus $T$ is the standard error of the y-intercept, expressed as $0.69\sqrt{\dfrac{\sum_{i=1}^{n}x_i^2 \sum_{i=1}^{n}(y_i - \hat{y}_i)^2}{n(n-2)\sum_{i=1}^{n}(x_i - \bar{x})^2}}$, where $\hat{y}_i$ is the corresponding value predicted by linear regression on $n$ sets of data points $(x_i, y_i)$ and $\bar{x}$ is the mean of $x_i$. Critical current measurements were performed using a 60 Hz alternating current at temperatures ranging from 2 K to 12 K (Fig. S23). Voltage was recorded as the current was swept from 0 until a jump was observed.

## 2 Supplementary Text

2.1 Elemental composition of NbN-type materials

Elemental composition has been found to influence the superconducting behavior of NbN. For example, $T_c$ and $B_{c2}$ depend on the ratio between C and N in the bulk Nb–N–C system, with the highest $T_c$ observed at around 30 % C/Nb ratio[20,24,26]. Since our 3D printed NbN-type materials were derived by nitridation of the parent oxide samples (calcined from BCP-niobia sol hybrids) in ammonia followed by reduction and annealing under carburizing gas that contained hydrogen and methane, carbon was expected to be incorporated while some of the oxygen was expected to remain.

An accurate quantitative assessment of light elements such as C, N, and O can be challenging, especially when these elements exist as contaminants from the environment. XPS survey scans showed that the C atomic percentage dropped substantially after etching (*e.g*., by more than three quarters from ≈ 20 % to ≈ 5 % after etching NbN-type samples treated to 750 °C under carburizing gas, see Section 1.5.7; Fig. S9 and Table S1), with both values (*i.e*., before and after etching) in turn being substantially higher than those of the parent oxides. The significant reduction of carbon upon etching suggests that it accumulates on the surface of these high-surface-area materials, likely due to deposition from methane particularly at temperatures above 800 °C (Table S1). Therefore, we subsequently focused our attention on XPS of etched samples. It should be emphasized that etching of the samples is expected to be depth dependent rather than homogeneous across the entire sample. Therefore, results on etched samples provide general trends rather than reliable absolute values.

High-resolution scans of the C 1s peak (Fig. S10) helped elucidate the form of carbon in the NbN-type materials as carbon could exist in various forms. One possibility is as carbide in NbC, which is also a superconductor and forms a binary solution with NbN[26]. The carbide content increased from ≈ 10 % to ≈ 20 % as the temperature to which the sample was treated under carburizing gas increased from 750 °C to 950 °C (Table S2). At the same time, relative to Nb, the N content stayed approximately the same while the O content dropped by around two thirds (Table S1). This suggested that higher temperatures during the second heat treatment step in carburizing gas increased incorporation of C into the rock salt lattice, displacing O, while the N content stayed roughly the same. Half of the C content of these samples was accounted for by C–C bonded C, suggesting substantial formation of elemental carbon during heat treatment in carburizing gas. As a result of limitations of lab-based XPS, in particular for porous samples studied here, the data shown in Table S2 should be used for analyzing qualitative trends rather than relied upon for absolute values.

From these results, the materials generated through thermal treatments under different environments (ammonia and carburizing gases) are essentially niobium oxycarbonitrides. Both WAXS (Fig. S8b) and TEM (Fig. S11b) data sets were consistent with rock salt structures of both NbC and NbN that are difficult to distinguish due to the very close lattice parameters (4.39 Å for NbN versus 4.43 Å NbC)[43-45]. Based on all these considerations, in the text we have consistently referred to these complex superconducting materials compositions as "NbN-type materials".

2.2 $B_{c2}(0)$ of NbN-type materials heat treated at 750 °C and structure-directed by ISO

Since the high critical field values of our NbN-type SCs exceeded the capability of the magnet inside the PPMS (limited to 14 T, *vide supra*), $B_{c2}(0)$ values were extrapolated from the slope of the linear regression curve, $\frac{dB_{c2}}{dT}$, near $T_c$ through $B_{c2}(0) = 0.69T_c \left.\frac{dB_{c2}}{dT}\right|_{T=T_c}$, according to the Werthamer–Helfand–Hohenberg (WHH) theory[29]. The error bars associated with this extrapolation, depicted in Fig. 4d, represent errors arising from linear regression. This is different from the error bars in Fig. 4f, which represent the standard deviations of results of multiple measurements. Having a high $T_c$ that is only weakly suppressed by increasing fields is conducive to a high $B_{c2}(0)$. The NbN-type material heat treated to 750 °C in carburizing gas before optimization had a low $T_c$ of around 7 K (see Fig. 3f in main text), limiting the extrapolated $B_{c2}(0)$ value (Fig. S24). On the other hand, the $T_c$ of NbN-type materials structure-directed by ISO did not linearly depend on the external magnetic field. Similar to earlier findings[46], the residual for linear regression was smallest when $B$ was plotted against $T^4$, which did not yield a high $B_{c2}(0)$ despite a high $T_c$ (Fig. S25).

2.3 Correlation between wall thickness and molar mass of PEO block

In Fig. 4f, the wall thickness, $d$, was plotted on the top horizontal axis, which was assumed to be proportional to the molar mass of the PEO block ($M_{n,\text{PEO}}$) raised to the power of two thirds. Although the periodic spacing of BCP SA directed mesophases has been found both theoretically and experimentally to be proportional to molar mass to the power of two thirds[36,47-49], the proportionality constant, $k$, was based on the equilibrium morphologies of pure BCPs processed either in the melt or from solution. In contrast, in the current case the dimension of concern was the wall thickness of the inorganic NbN-type materials after heat treatment and associated substantial shrinkage from combustion of the BCP structure directing agents and densification of the sol-gel derived inorganic material. Provided the relationship between $d$ and $M_{n,\text{PEO}}$ still followed the relation: $d = kM_{n,\text{PEO}}^{2/3}$, $k$ could be first calculated based on the average $d$ value determined from measuring the pore sizes in SEM micrographs of NbN-type mesostructures directed by F127. This provided $d = 4.6$ nm for the sample treated to 950 °C in carburizing gas. Putting $M_{n,\text{PEO}}$ of F127 in units of kg/mol, $k$ was found to be 1.08 nm.

The thickness could be alternatively obtained by subtracting the pore size, $W$, determined by gas sorption from the correlation length between adjacent pore centers, $d_0$, determined by SAXS. $d_0$ can be calculated from $q^*$ of the primary peak position via $d_0 = \frac{4\pi}{\sqrt{3}q^*}$. Note that $d$ calculated in this way represents the thinnest part in a perfect hexagonal structure with cylindrical pores. However, the choice of different models used for generating pore size distributions resulted in large variations of $W$. Using the nitride sample treated to 750 °C under carburizing gas as an example, $W$ was found to be 9.0 nm from the NLDFT model (Fig. S4b) versus 3.9 nm from the BJH model (Fig. 3d) for the NbN-type material heat treated to 750 °C in carburizing gas. With a spacing of $d_0 = 11.7$ nm (Fig. 3a) determined from SAXS results, this led to a wall thickness possibly ranging from 2.7 nm to 7.8 nm, a range too wide to offer any meaningful characterization of the true size. Therefore, we resorted to direct measurements of the wall thickness in the SEM micrograph of the nitride sample treated to 950 °C under carburizing gas to calculate $k$. The validity of our approach was corroborated by the calculated wall thickness of 13.4 nm for NbN-type materials structured directed by ISO with $M_{n,\text{PEO}} = 43.7$ kg/mol, close to 15.2 nm measured from SEM (Fig. S17).

2.4 3D printing BCP-titania sol ink and resulting oxide and nitride materials

During DIW, a slower printing speed of 100 mm/min was used, likely due to the slower coagulation kinetics of the F127-titania ink in hexane. Calcination in air and subsequent heat treatment in ammonia yielded cubic oxide and nitride woodpiles, respectively (Fig. S5b,c). Another difference from the niobium system is the lower oxide crystallization temperature, which allowed us to prepare ordered crystalline $TiO_2$ with small mesopores directed by the low molar mass Pluronics BCPs. Depending on the calcination temperature in air, the oxide could be chosen to be either amorphous (heated to 300 °C, yellow curve in Fig. S6b) or crystalline (heated to 400 °C, orange curve in Fig. S6b).

Compared with its niobium counterpart, TiN had a greater propensity for crystal coarsening, as reflected by the diminished SAXS peak signal (Fig. S6a) and the sharper WAXS peak profile (Fig. S6b). Nitrogen sorption data still suggested a narrow pore size distribution (Fig. S6c,d), with specific surface areas of 368 m$^2$/g and 233 m$^2$/g for the oxides calcined at 300 °C and 400 °C, and 113 m$^2$/g for the nitride heat treated to 750 °C in ammonia, respectively. This TiN sample was, however, not superconducting. Only by heating to a temperature at or above 900 °C did the sample become superconducting, with a $T_c$ between 3 K and 4 K as determined from electrical transport and VSM measurements (Fig. S6e,f). Despite greater challenges associated with controlling crystal growth in this materials system, preliminary non-optimized thermal processing protocols suggested successful access to first 3D printed mesoporous transition metal compounds, including superconducting nitrides. Further fine-tuning will be required, however, to achieve structure control levels comparable to the NbN-type material system.

## 3 Supplementary Figures

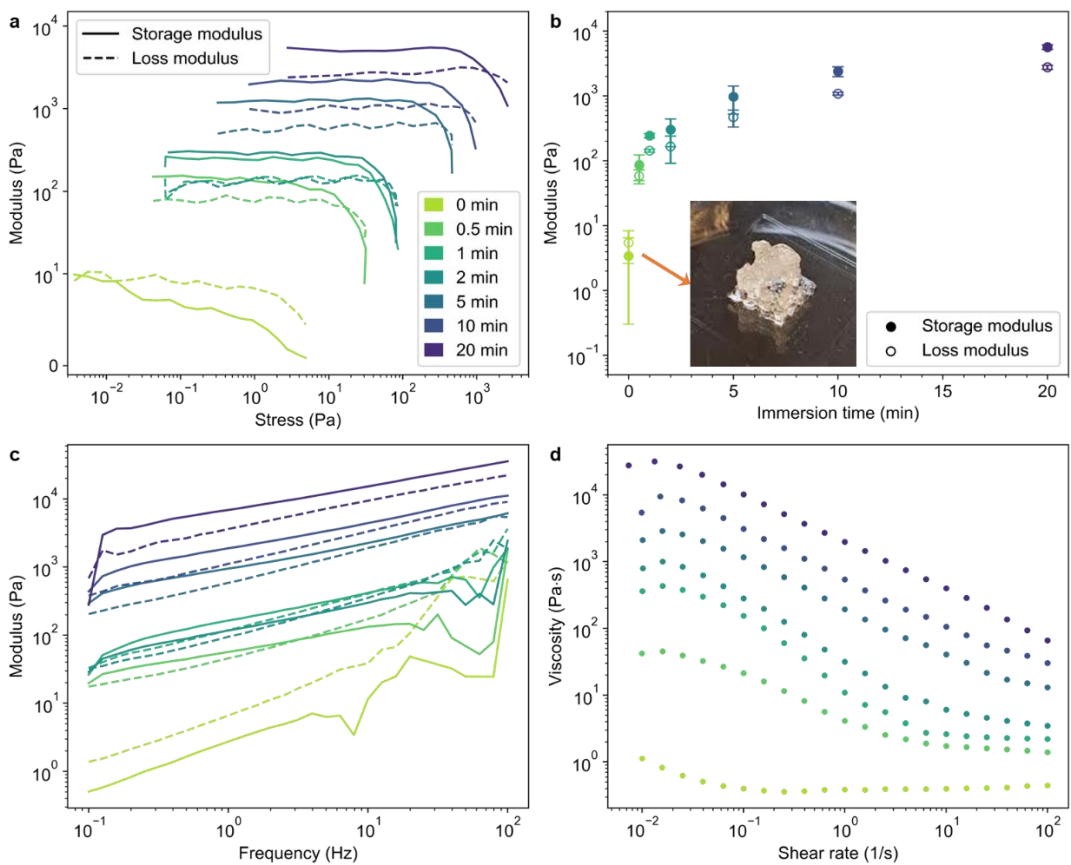

**Fig. S1.** Rheological characterization of inks immersed in hexane for various durations. (a) Plot of storage and loss moduli as a function of stress as the strain was swept from 0.01 % to 100 % at the frequency of 1 Hz in the oscillatory mode. (b) Plot of storage and loss moduli averaged over the plateau region in (a) as a function of immersion time. Inset: Photo of a woodpile structure showing poor structure retention after printing directly in air without being immersed in hexane. (c) Plot of storage and loss moduli as a function of frequency at 1 % strain. (d) Plot of viscosity as a function of shear rate in the flow mode. The same color scheme regarding immersion time in (a) applies to (b-d). The same line style for storage and loss moduli is used in (a,c).

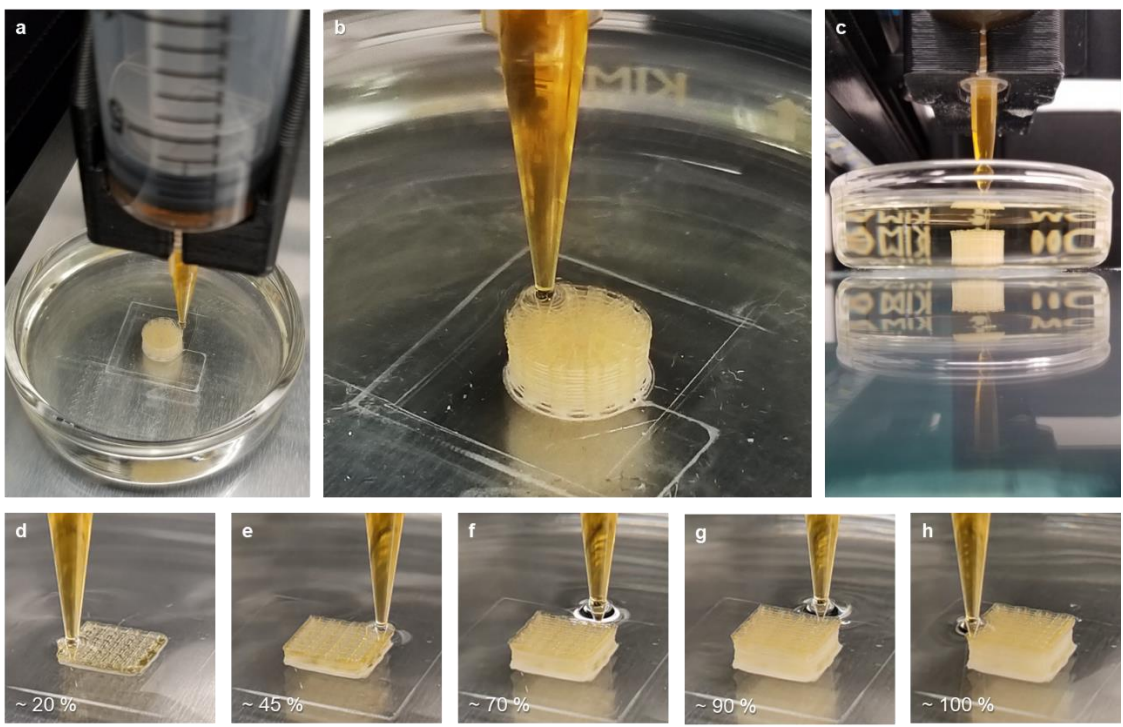

**Fig. S2.** (a-c) Photos taken from different angles during printing of a cylindrical woodpile. The Sudan I dye has a higher solubility in hexane than in alcohol, so the hexane bath became orange as the dye diffused out of the acidic alcoholic ink. (d-h) Photos taken at progressing completion levels during printing of a cubic woodpile.

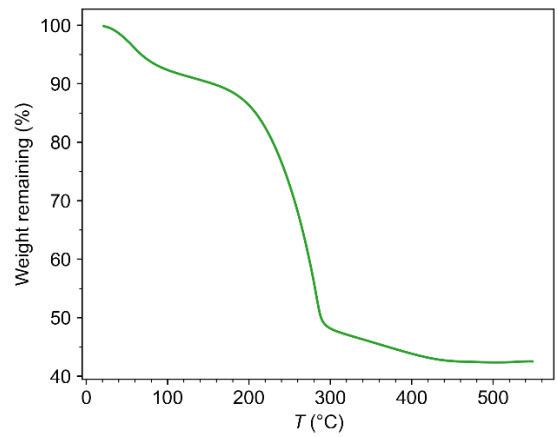

**Fig. S3.** Thermogravimetric analysis plot of remaining weight percentage of a 3D printed F127-noibium hybrid calcined in air at a ramp rate of 4 °C/min.

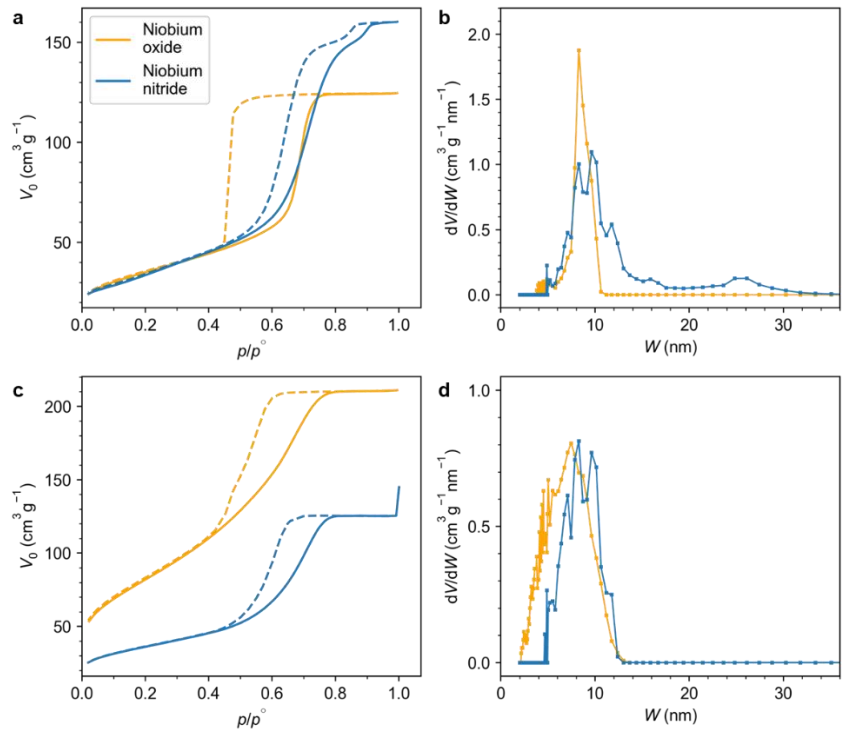

**Fig. S4.** Nitrogen adsorption and desorption curves of mesoporous oxide and nitride-type (a) woodpile samples shown in Fig. 3c and (c) helical samples shown in Fig. 5j. Corresponding pore size distributions for (b) the woodpile samples and (d) the helical samples, derived from the NLDFT model as opposed to the Barrett–Joyner–Halenda (BJH) model-based results shown in Fig. 3d and Fig. 5k in the main text. All oxides were derived from calcination in air to 450 °C and the nitrides were derived by heat treating the oxide first under ammonia to 550 °C and then under carburizing gas to 750 °C. The same color scheme in (a) applies to (b-d).

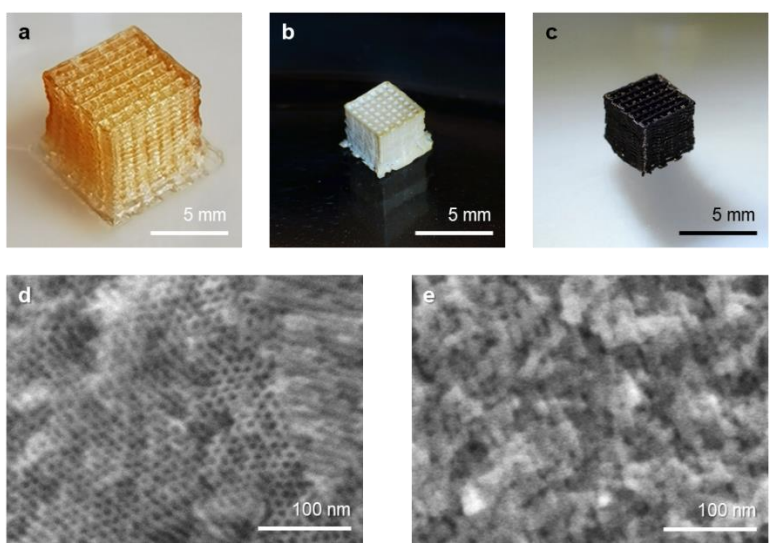

**Fig. S5.** 3D printed structures derived from BCP-titania sol. (a-c) Photos of as-printed cubic woodpile of Pluronic F127-titania sol hybrid (a) and resulting oxide (b) and nitride (c) woodpiles. (d, e) SEM micrographs of mesostructures of oxide (d) and nitride (e).

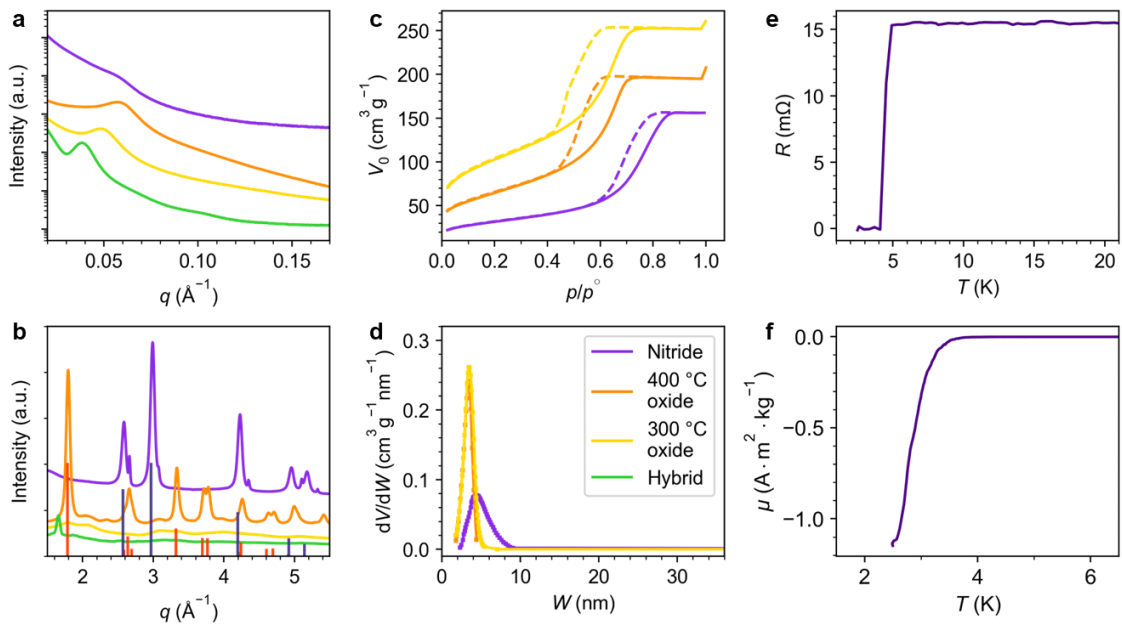

**Fig. S6.** (a) SAXS and (b) WAXS profiles for 3D printed Pluronic F127-titania sol hybrid and resulting oxides and nitride structures. Red and navy ticks in (b) represent relative intensities and $q$ positions of peaks of anatase (PDF #98-000-0081) and titanium nitride (TiN) (PDF #01-087-0632), respectively. The nitride sample was characterized after electrical transport measurements, so the silver paint used to contact the sample gave rise to additional WAXS peaks that are not attributable to nitride materials. (c) Nitrogen adsorption (solid line) and desorption (dashed line) curves of mesoporous titanium oxides and nitride. (d) Corresponding pore size distributions derived from the BJH model. The yellow and orange curves superimpose so well that they are difficult to distinguish. (c-d) share the same legend shown in (d). (e) Plot of electrical resistance ($R$) as a function of temperature ($T$) of TiN sample heat treated in ammonia to 900 °C. (f) Plot of magnetic moment normalized by sample mass ($\mu$) as a function of $T$ of TiN sample heat treated in ammonia to 950 °C.

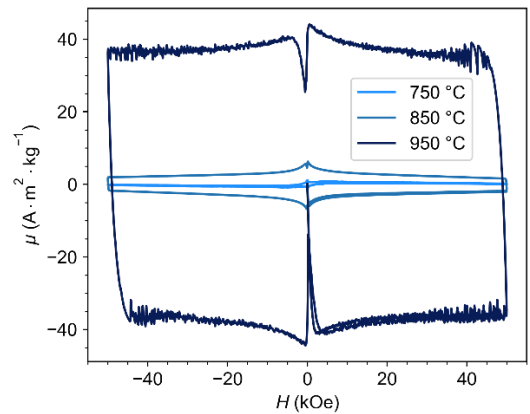

**Fig. S7.** Plot of magnetic moment normalized by mass ($\mu$) measured at 1.8 K as a function of external field (*H*) for NbN-type materials heat treated to different temperatures in the final heating step under carburizing gas (as indicated). F127 was the structure directing BCP.

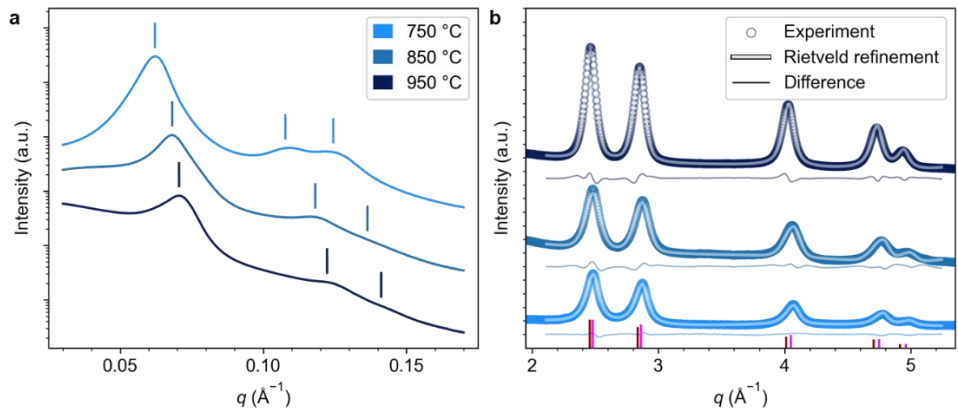

**Fig. S8.** (a) SAXS and (b) WAXS profiles for F127-directed NbN-type materials heat treated to different temperatures in carburizing gas. Ticks above curves in (a) denote relative $q$ positions (1:$\sqrt{3}$:2) of Bragg reflections for hexagonally ordered structures. The spacings, $d_0$, as determined from the primary peak positions decrease from 11.7 nm for 750 °C, through 10.6 nm for 850 °C, to 10.3 nm for 950 °C. The dark red and magenta ticks in (b) represent expected relative intensities and corresponding $q$ positions of peaks of rock salt NbN (PDF #01-089-5007) and NbC (PDF #03-065-8781). Rietveld refinement results (white curve of circles through data) and corresponding difference (below each curve) from the experimental data (circles) are also plotted. The same color scheme in (a) applies to (b).

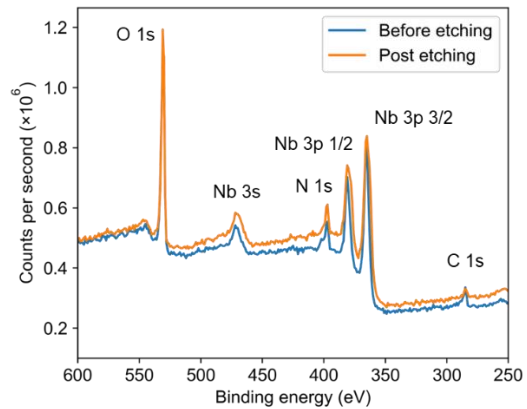

**Fig. S9.** XPS survey scan spectra of a 3D printed NbN-type sample treated to 550 °C under ammonia and to 750 °C in the carburizing gas before and after etching. See Table S1 for elemental composition analysis.

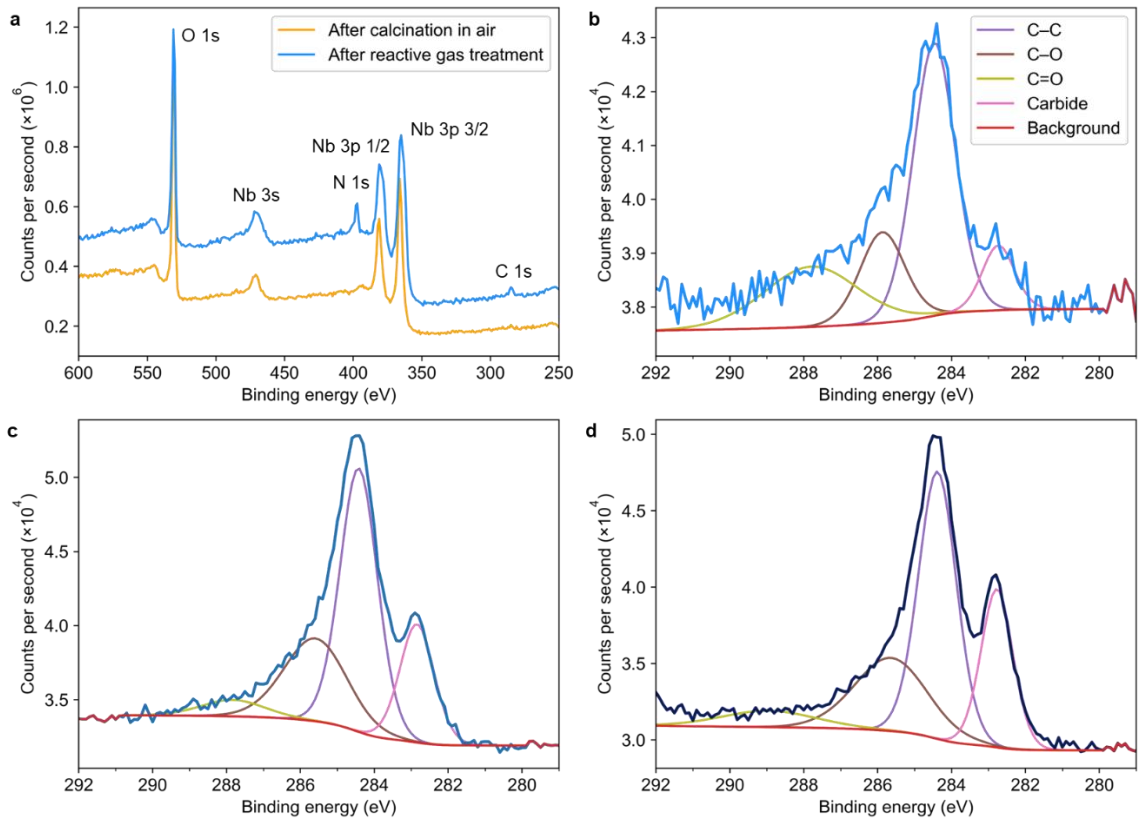

**Fig. S10.** (a) XPS spectra for 3D printed niobium oxide after calcination in air and niobium nitride heat treated to 550 °C under ammonia and then to 750 °C in carburizing gas. High-resolution scans of C 1s peak for samples heat treated to 550 °C under ammonia and then to (b) 750 °C, (c) 850 °C, and (d) 950 °C fitted with different chemical bonding components of carbon. The same color scheme in (b) applies to (c,d). See Table S2 for chemical bonding information analysis.

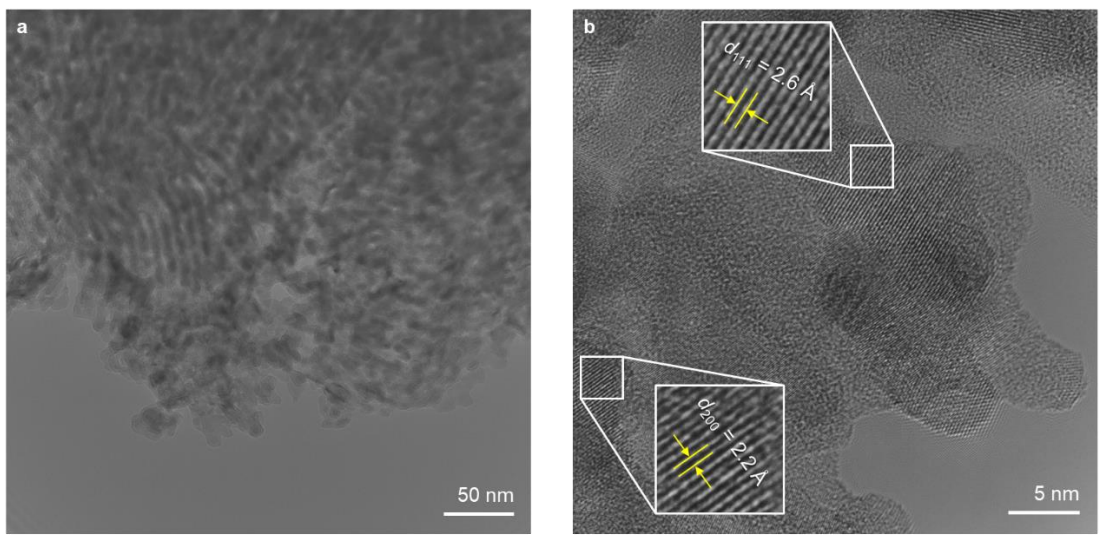

**Fig. S11**. (a) TEM image at a lower magnification consistent with self-assembled cylindrical mesostructures (see projections of cylinders laying down). The average distance of 8.8 nm across parallel cylinders matches the $d_0$ spacing determined through SAXS in Fig. S8a. (b) High-resolution TEM image showing a granular crystal structure after heat treatment in carburizing gas to 950 °C. Enlarged areas show interplanar distances of (111) and (200) planes of the rock salt structure, consistent with unit cell parameters derived from WAXS.

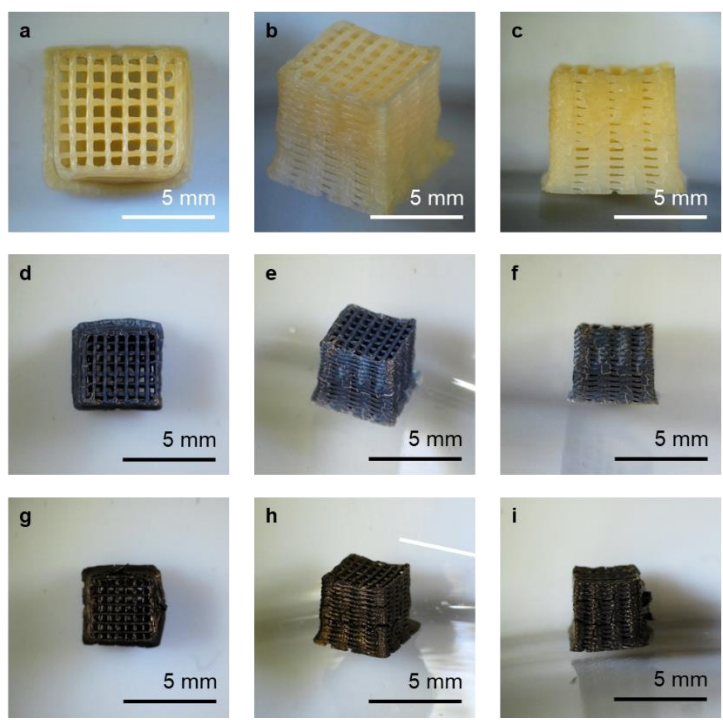

**Fig. S12.** Photos from different angles of (a-c) hybrid, (d-f) oxide, and (g-i) nitride woodpiles directed by Pluronics-family BCP F108. The oxide was derived from calcination in air to 450 °C and the nitride was derived by heat treating the oxide first under ammonia to 550 °C and then under carburizing gas to 750 °C. The black color of the oxide woodpile is a result of minor amounts of residual carbon.

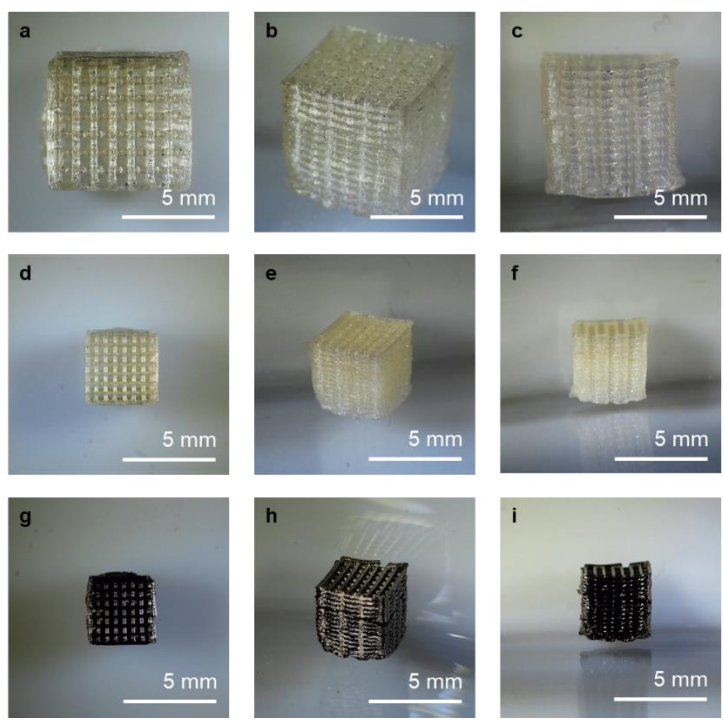

**Fig. S13.** Photos from different angles of (a-c) hybrid, (d-f) oxide, and (g-i) nitride woodpiles directed by Pluronics-family BCP P123. The oxide was derived from calcination in air to 450 °C and the nitride was derived by heat treating the oxide first under ammonia to 550 °C and then under carburizing gas to 750 °C.

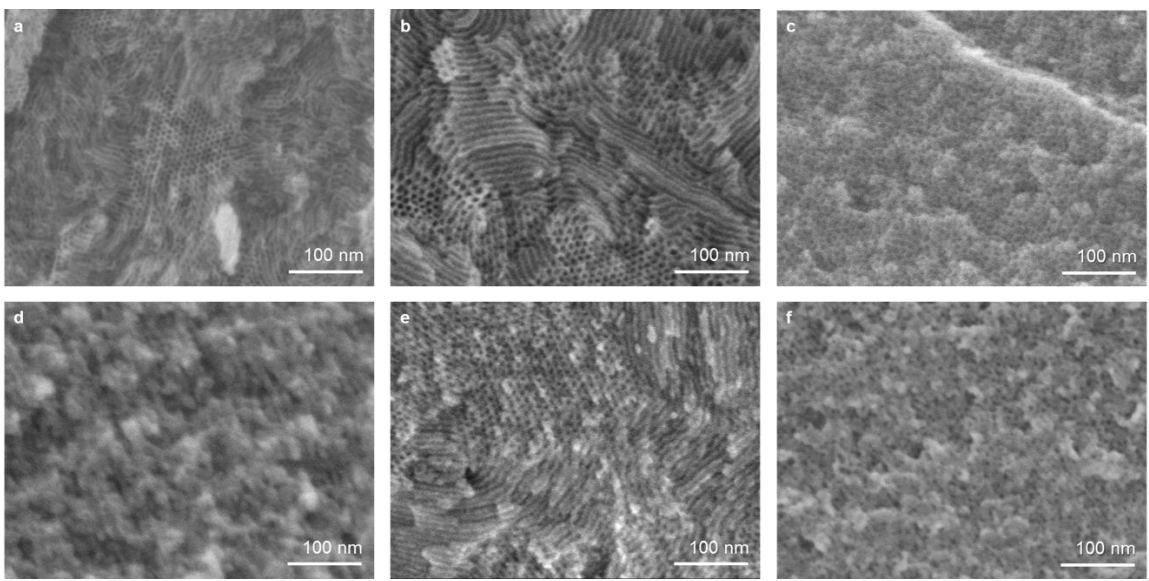

**Fig. S14.** SEM micrographs of 3D printed and heat treated (a-c) niobium oxide and (d-f) NbN-type materials structure-directed by different Pluronic BCPs: (a,d) P123, (b,e) F127, and (c,f) F108. All nitride samples were first heat treated to 550 °C under ammonia and then to 950 °C under carburizing gas.

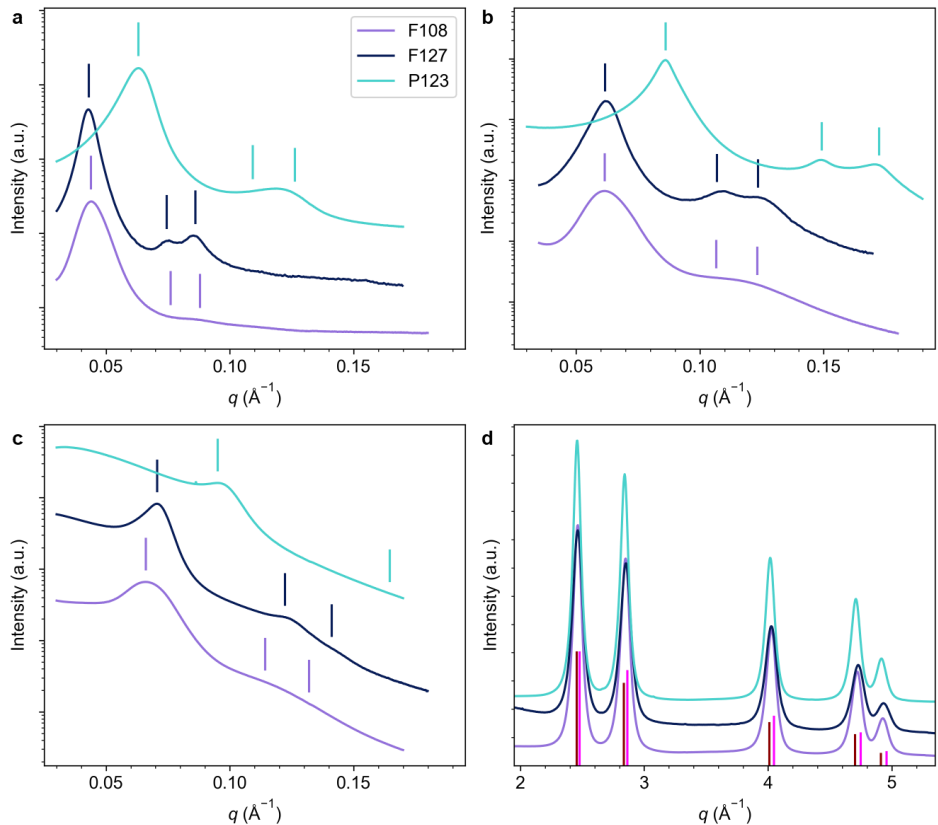

**Fig. S15.** (a-c) SAXS and (d) WAXS profiles for 3D printed (a) hybrid, heat treated (b) niobium oxide, and (c) NbN-type materials structure-directed by different Pluronic BCPs. All nitride samples were first heat treated to 550 °C under ammonia and then to 950 °C under carburizing gas. Ticks above curves in (a-c) denote relative $q$ positions (1:$\sqrt{3}$:2) of Bragg reflections for hexagonally ordered structures. The dark red and magenta ticks in (d) represent expected relative intensities and corresponding $q$ positions of peaks of rock salt NbN (PDF #01-089-5007) and NbC (PDF #03-065-8781), respectively. The same color scheme in (a) applies to (b-d).

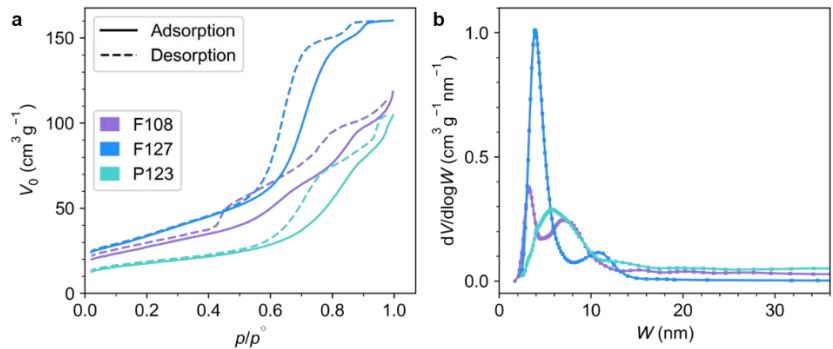

**Fig. S16.** (a) Nitrogen adsorption and desorption curves of NbN-type materials structure-directed by different Pluronic BCPs as indicated. All samples were treated to 750 °C under carburizing gas in the final heating step. (b) Corresponding pore size distributions derived from the Barrett–Joyner–Halenda (BJH) model. Results shown for F127 are identical to those shown in Fig. 3c,d of the main text. The same color scheme in (a) applies to (b).

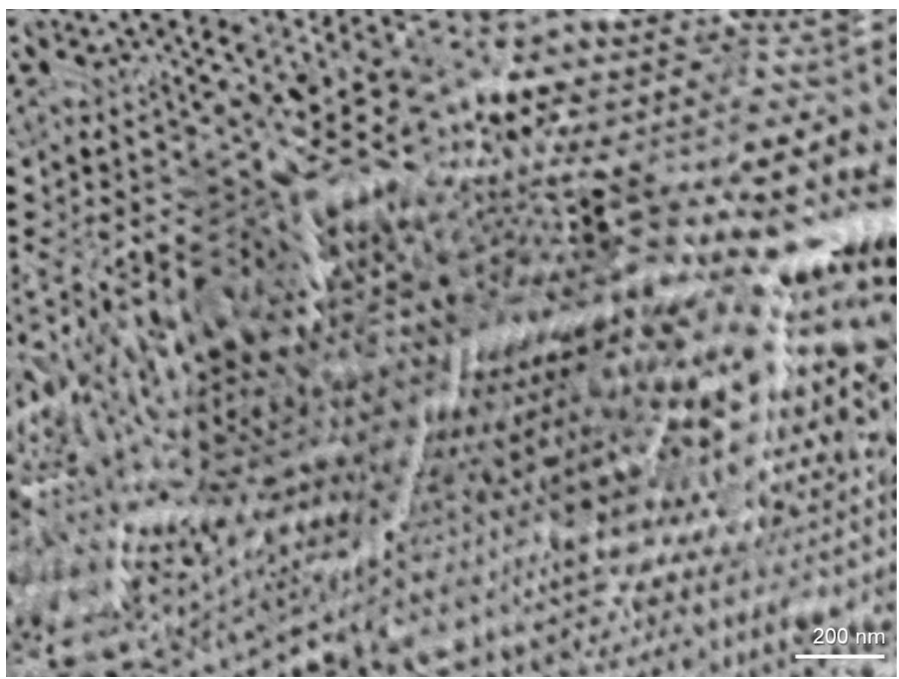

**Fig. S17.** SEM micrograph of a hexagonally structured NbN-type superconductor directed by poly(isoprene-*b*-styrene-*b*-ethylene oxide) with a molar mass of 88.3 kg/mol, *i.e.*, substantially larger than that of the Pluronics-type BCPs used in this study.

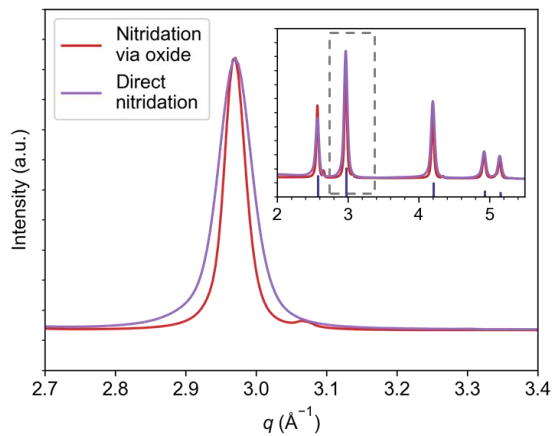

**Fig. S18.** Comparing WAXS results of TiN samples heat treated using different protocols. Nitridation via oxide refers to the original two-step protocol where the hybrid sample was first calcined in air at 300 °C to form the amorphous oxide and then heat treated in ammonia in a separate step. Direct nitridation refers to the refined one-step protocol where the hybrid sample was directly heat treated in ammonia. Both samples were treated to the final temperature of 900 °C in ammonia. The small hump at around 3.07 Å$^{-1}$ is due to the silver paint used to contact the sample for electrical transport measurements. The inset shows the full WAXS profiles with the peak shown in the main panel highlighted by the dashed rectangle. The ticks at the bottom represent expected $q$ positions and relative intensities of peaks of titanium nitride (TiN) (PDF #01-087-0632).

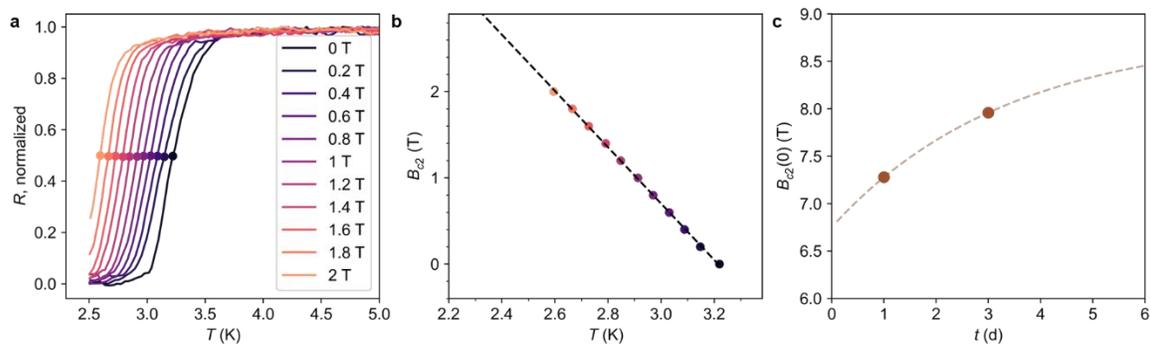

**Fig. S19.** Plots of (a) normalized $R$ of a TiN sample, structure-directed by F127 and heat treated directly to 900 °C in ammonia, as a function of $T$ at varying applied fields, and (b) upper critical field, $B_{c2}$, as a function of $T$ near $T_c$ from data in (a). (c) $B_{c2}(0)$ changing with aging time ($t$) in days in the ambient. In contrast to Fig. 4d, the dashed curve here is a visual guide only.

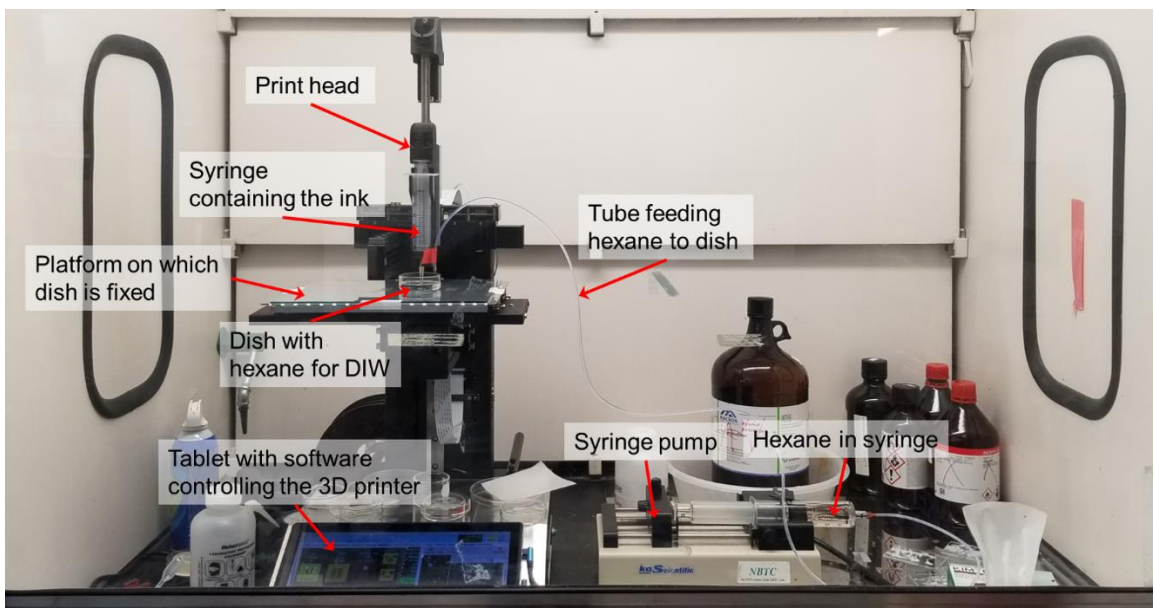

**Fig. S20.** Photo showing the entire DIW setup inside a fume hood.

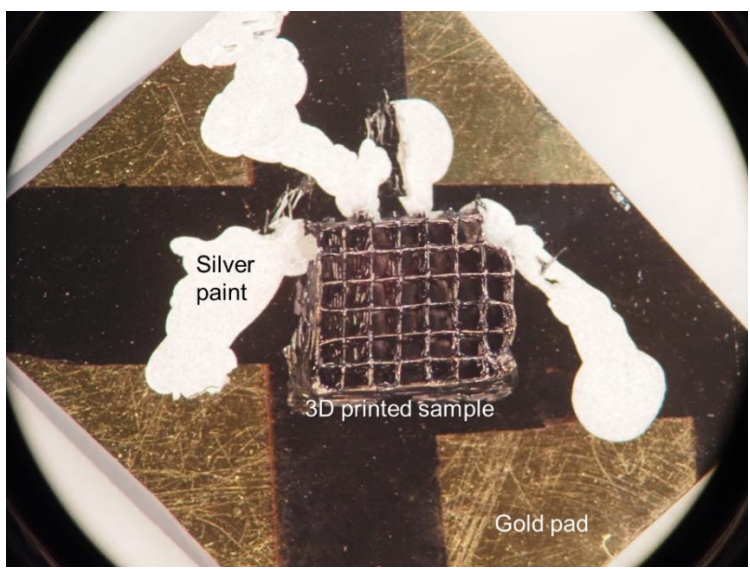

**Fig. S21.** Optical microscopy image of a NbN-type woodpile on a chip connected to gold pads through silver paint. The gold pads are subsequently bonded through wires to a resistivity puck to be used in electrical transport measurements in the PPMS unit. In most instances, smaller sample pieces as compared to the one shown here were broken from a whole woodpile and then measured.

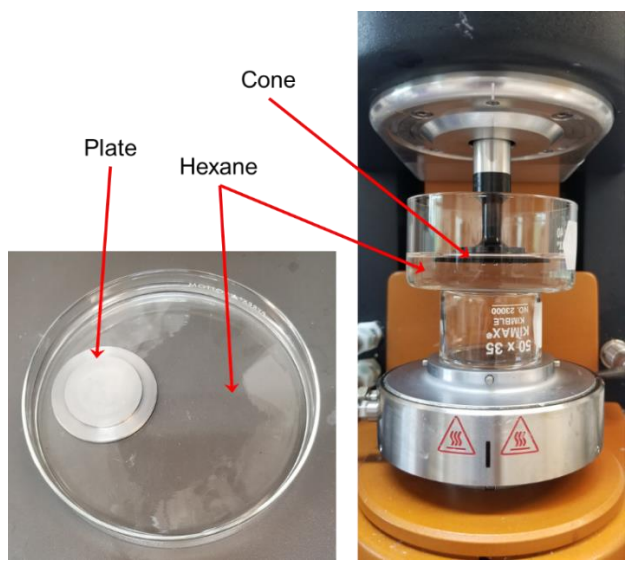

**Fig. S22.** Photo showing cone (black, right image) and plate (silver, lower left image) immersed in hexane for rheological measurements.

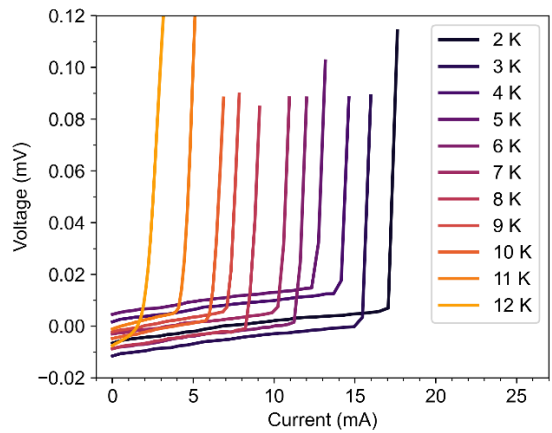

**Fig. S23.** Plot of voltage as a function of current for a NbN-type sample at different temperatures. The F127-directed NbN-type material used was heat treated to 575 °C under ammonia and then to 950 °C under carburizing gas.

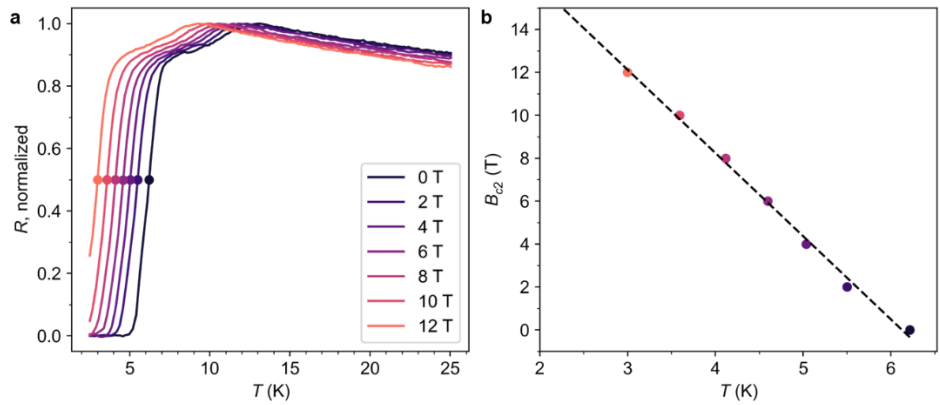

**Fig. S24.** Plots of (a) normalized $R$ of a NbN-type sample structure-directed by F127 and heat treated to 750 °C in carburizing gas as a function of $T$ at varying applied fields, and (b) upper critical field, $B_{c2}$, as a function of $T$ near $T_c$.

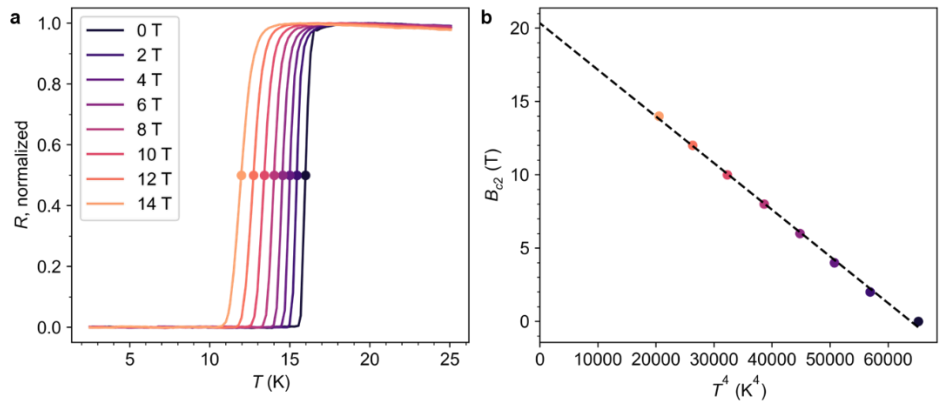

**Fig. S25.** Plots of (a) normalized $R$ of a NbN-type sample structure-directed by large molar mass BCP ISO as a function of $T$ at varying applied fields, and (b) upper critical field, $B_{c2}$, as a function of $T^4$ to extrapolate the $B_{c2}$ value at $T = 0$, $B_{c2}(0)$.

## 4 Tables

**Table S1.** Elemental composition of 3D printed oxide and NbN-type materials*, the latter heat treated to 550 °C under ammonia and then to higher temperatures (as indicated) under carburizing gas.

|  | Before etching | | | | Post etching | | | |
|---|---|---|---|---|---|---|---|---|
|  | Nb | O | N | C | Nb | O | N | C |
| Oxide | 27.4 | 65.2 | 0 | 7.4 | 32.1 | 66.6 | 0 | 1.3 |
| 750 °C nitride | 24.0 | 44.2 | 10.5 | 21.3 | 34.9 | 48.2 | 11.8 | 5.1 |
| 850 °C nitride | 21.3 | 17.1 | 14.5 | 47.1 | 28.4 | 30.0 | 10.8 | 30.8 |
| 950 °C nitride | 23.4 | 22.4 | 13.2 | 41.0 | 30.6 | 17.8 | 11.6 | 40.0 |

* Atom % determined by XPS survey scans.

**Table S2.** Chemical bonding information of carbon in 3D printed NbN-type materials* heat treated to 550 °C under ammonia and then to higher temperatures (as indicated) under carburizing gas.

|  | carbide | C–O | C=O | C–C |
|---|---|---|---|---|
| 750 °C nitride | 9.0 | 17.3 | 24.4 | 49.3 |
| 850 °C nitride | 20.1 | 25.7 | 5.1 | 49.0 |
| 950 °C nitride | 22.9 | 24.2 | 7.0 | 46.0 |

* Atom % of total carbon determined by XPS high-resolution scans of the C 1s peak for samples after etching.

**Table S3**. Structural characteristics of NbN-type samples at various stages of processing.

| Structure-directing BCP | Materials | Center-to-center distance[§] $d_0$ (nm) | Unit cell size[#] $a$ (Å) | Coherently scattering domain size[$] $\tau$ (nm) | Pore size[^] $W$ (nm) | Wall thickness[‖] $d$ (nm) |
|---|---|---|---|---|---|---|
| F127 | Hybrid | 16.9 | – | – | – | – |
| | Oxide | 11.8 | – | – | 8.3/3.8 | 3.5/8.0 |
| | 750 °C nitride[†] | 11.7 | 4.352 | 4.0 | 9.0/3.9 | 2.7/7.8 |
| | 850 °C nitride[†] | 10.6 | 4.356 | 4.0 | – | – |
| | 950 °C nitride[†] | 10.3 | 4.400 | 4.7 | – | 4.6[Δ] |
| ISO | 1000 °C nitride[‡] | 39.7[⊥] | – | – | – | 15.4[Δ] |

† Treated first under ammonia at 550 °C and then under carburizing gas at indicated temperatures.
‡ Treated first under ammonia at 700 °C.
§ Determined by the scattering vector of the primary Bragg reflection peak, $q^*$, in SAXS through $d_0 = \frac{4\pi}{\sqrt{3}q^*}$, unless otherwise noted.
⊥ Measured on SEM micrographs.
# Determined by WAXS and whole pattern fitting using the JADE software.
$ Determined by WAXS and Scherrer analysis using the JADE software.
^ Determined by nitrogen sorption. Numbers before and after the slash are pore sizes derived from the NLDFT and BJH models, respectively.
‖ Calculated by $d_0 - W$, unless otherwise noted. Numbers before and after the slash are calculation results using $W$ derived from the NLDFT and BJH models, respectively.
Δ Measured on SEM micrographs.

**Table S4.** Structural characteristics of 3D printed NbN-type materials structure-directed by different Pluronic BCPs.

| Structure-directing BCP | Materials | Center-to-center distance[§] $d_0$ (nm) | Unit cell size[#] $a$ (Å) | Coherently scattering domain size[$] $\tau$ (nm) | Pore size[^] $W$ (nm) | BET specific surface area[¤] $S$ (m$^2$/g) |
|---|---|---|---|---|---|---|
| P123 | Hybrid | 11.5 | – | – | – | – |
| | Oxide | 8.4 | – | – | 2.7 | – |
| | Nitride[¶] | 7.6 | 4.436 | 6.7 | 5.8 | 62.5 |
| F127 | Hybrid | 16.9 | – | – | – | – |
| | Oxide | 11.8 | – | – | 3.8 | – |
| | Nitride[¶] | 10.3 | 4.400 | 4.7 | 3.9 | 119.7 |
| F108 | Hybrid | 16.5 | – | – | – | – |
| | Oxide | 11.8 | – | – | 3.3 | – |
| | Nitride[¶] | 11.0 | 4.422 | 6.1 | 3.3 | 96.8 |

§ Determined by the scattering vector of the primary Bragg reflection peak, $q^*$, in SAXS through $d_0 = \dfrac{4\pi}{\sqrt{3}q^*}$, unless otherwise noted.

\# Determined by WAXS and whole pattern fitting using the JADE software.

$ Determined by WAXS and Scherrer analysis using the JADE software.

^ Determined by nitrogen sorption using the BJH model.

¤ Determined by nitrogen sorption.

¶ Nitride results are from non-optimized samples. $d_0$, $a$, and $\tau$ values are obtained on nitride samples treated to 950 °C under carburizing gas. $W$ and $S$ values are obtained on nitride samples treated to 750 °C under carburizing gas.

**Movie S1.** Direct ink writing F127-niobia sol ink in hexane

**Movie S2.** Embedded printing F127-niobia sol ink in F127 "gel"

**Movie S3.** Compressing & releasing helix in ethanol (part 1)

**Movie S4.** Compressing & releasing helix in ethanol (part 2)

**Movie S5.** Fishing out helix from ethanol with tweezers after rinsing

**Glossary**

| | |
|---|---|
| 2D | two-dimensional |
| 3D | three-dimensional |
| $\beta$ | stretching exponent (in Kohlrausch–Williams–Watts function) |
| $\mu$ | magnetic moment (normalized by mass) |
| $\xi$ | (Ginsburg–Landau) coherence length (of a superconductor) |
| $\xi_0$ | coherence length (of a superconductor) in the clean limit |
| $\xi(T)$ | (Ginsburg–Landau) coherence length (of a superconductor) at temperature $T$ |
| $\tau$ | coherently scattering domain size |
| $\Phi_0$ | magnetic flux quantum |
| $a$ | unit cell size |
| AcOH | acetic acid |
| a.u. | arbitrary unit |
| $B_{c2}$ | upper critical field (of a type-II superconductor) |
| $B_{c2}(0)$ | upper critical field (of a type-II superconductor) at absolute zero temperature |
| BCP | block copolymer |
| BET | Brunauer–Emmett–Teller (theory for measuring surface area) |
| BJH | Barrett–Joyner–Halenda (model for obtaining pore size distribution) |
| CHESS | Cornell High Energy Synchrotron Source |
| $d$ | wall thickness |
| $d_0$ | characteristic length between adjacent pore centers |
| $Đ$ | dispersity (of molecular weight) |
| DIW | direct ink writing |
| $e$ | elementary charge |
| EISA | evaporation-induced self-assembly |
| EtOH | ethanol |
| FC | field cooling |
| FMB | Functional Materials Beamline |
| GPC | gel permeation chromatography |
| *h* | Planck constant |
| *H* | applied external magnetic field strength |
| HRTEM | high-resolution transmission electron microscopy |
| ISO | poly(isoprene-*block*-styrene-*block*-ethylene oxide) |
| KWW | Kohlrausch–Williams–Watts (function) |
| $l_{tr}$ | transport mean free path |
| *m* | mass |
| $M_n$ | number-average molar mass |
| $M_{n,PEO}$ | number-average molar mass of polyethylene |
| NLDFT | non-local density functional theory |
| NMR | nuclear magnetic resonance |
| NSLS-II | National Synchrotron Light Source II |
| *p* | pressure (of adsorbate in gas sorption) |
| *p*° | saturation pressure (of pure adsorbate in gas sorption) |
| PDF | powder diffraction file |
| PDI | polydispersity index |
| PEO | polyethylene oxide |
| PI | polyisoprene |
| PPMS | Physical Property Measurement System |
| PPO | polypropylene oxide |
| PS | polystyrene |
| *q* | (magnitude of) scattering vector |
| $q^*$ | (magnitude of) scattering vector of the primary peak |
| *R* | electrical resistance |
| $S$ | BET specific surface area |
| SA | self-assembly |
| SAXS | small-angle X-ray scattering |
| SC | superconductor |
| SEM | scanning electron microscopy |
| SMI | Soft Matter Interfaces (beamline) |
| $t$ | time |
| $T$ | temperature |
| TEM | transmission electron microscopy |
| TGA | thermogravimetric analysis |
| THF | tetrahydrofuran |
| $T_m$ | melting point |
| $V$ | volume |
| $V_0$ | volume of adsorbate adsorbed (in gas sorption) at standard temperature and pressure |
| VSM | vibrating sample magnetometry |
| $T_c$ | critical temperature, or superconducting transition temperature |
| $W$ | pore width |
| WAXS | wide-angle X-ray scattering |
| WHH | Werthamer–Helfand–Hohenberg (theory of superconductivity) |
| XPS | X-ray photoelectron spectroscopy |
| ZFC | zero-field cooling |

**Supplementary References**